\RequirePackage{snapshot}
\documentclass[11pt,onecolumn,twoside]{IEEEtran}

\usepackage{epsfig,bm,epsf,float,amsmath,amssymb,psfrag,subfigure,mathtools,url,color,algorithmic,algorithm,enumitem,cite}
\usepackage[section]{placeins} 
\usepackage{lipsum}

\usepackage[left=1in,right=1in,top=1in,bottom=1in]{geometry}

\usepackage[T1]{fontenc}
\usepackage[utf8]{inputenc}

\usepackage{dsfont}
\newcommand{\charfn}{\mathds{1}}

\newtheorem{theorem}{Theorem}
\newtheorem{proposition}{Proposition}
\newtheorem{lemma}{Lemma}

\newtheorem{definition}{Definition}

\newenvironment{remark}{\textit{Remark:}}

\providecommand{\eref}[1]{\eqref{eq:#1}}  
\providecommand{\cref}[1]{Chapter~\ref{chap:#1}}
\providecommand{\sref}[1]{Section~\ref{sec:#1}}
\providecommand{\fref}[1]{Figure~\ref{fig:#1}}

\providecommand{\NN}{\ensuremath{\mathcal{N}}}

\providecommand{\R}{\ensuremath{\mathbb{R}}}

\providecommand{\abs}[1]{\lvert#1\rvert}
\providecommand{\norm}[1]{\lVert#1\rVert}

\providecommand{\set}[1]{\left\{#1\right\}}

\providecommand{\bydef}{\overset{\text{def}}{=}}

\renewcommand{\vec}[1]{\boldsymbol{#1}}
\providecommand{\mat}[1]{\ensuremath{\boldsymbol{#1}}}

\DeclareMathOperator{\diag}{diag}
\providecommand{\calA}{\mathcal{A}}
\providecommand{\calB}{\mathcal{B}}

\providecommand{\calP}{\mathcal{P}}
\providecommand{\calH}{\mathcal{H}}

\providecommand{\calX}{\mathcal{X}}

\providecommand{\mA}{\mat{A}} \providecommand{\mB}{\mat{B}}
 
\providecommand{\mD}{\mat{D}}

 \providecommand{\mP}{\mat{P}}

\providecommand{\mX}{\mat{X}}
\providecommand{\mY}{\mat{Y}}

\providecommand{\va}{\vec{a}} \providecommand{\vb}{\vec{b}}

\providecommand{\vx}{\vec{x}} \providecommand{\vy}{\vec{y}}

\providecommand{\vone}{\vec{1}}
 \providecommand{\vv}{\vec{v}}

\providecommand{\var}{\operatorname{var}}

\newcommand{\EE}{\mathbb{E}}

\newcommand\hairspace{\kern .08333em }
\providecommand{\E}{\mathbb{E} \hairspace}

\providecommand{\lmax}{\lambda_{\max}}

\providecommand{\PF}{P_{\text{false\_alarm}}}
\providecommand{\PM}{P_{\text{miss}}}

\graphicspath{{figs/},{asy/}}

\begin{document}
\bstctlcite{NOURL}

\title{Optimal Detection of Random Walks on Graphs: Performance Analysis via \\Statistical Physics}
\author{Ameya~Agaskar%
            ,~\IEEEmembership{Student Member,~IEEE,} 
        and~Yue~M.~Lu,~\IEEEmembership{Senior Member,~IEEE}%
\thanks{The authors are with the School of Engineering and Applied Sciences, Harvard University, Cambridge, MA 02138, USA 
        (e-mail: \{aagaskar, yuelu\}@seas.harvard.edu).}
\thanks{A.~Agaskar is also with MIT Lincoln Laboratory. The Lincoln Laboratory portion of this work is sponsored by the Department of the Air Force under
            Contract FA8721-05-C-0002. Opinions, interpretations, conclusions and
            recommendations are those of the authors and are not necessarily
            endorsed by the United States Government.}
            \thanks{This paper was presented in part at the 2013 IEEE International Conference on Acoustics, Speech and Signal Processing and at the 2014 Asilomar Conference on Signals, Systems, and Computers.}
}

\markboth{}{Agaskar and Lu: Optimal Detection of Random Walks on Graphs}

\maketitle

\begin{abstract}
    We study the problem of detecting a random walk on a graph from a sequence of noisy measurements at every node.
    There are two hypotheses: either every observation is just meaningless zero-mean Gaussian noise, 
    or at each time step exactly one node has an elevated mean, with its location following a random walk on the graph
    over time.
    We want to exploit knowledge of the graph structure and random walk parameters (specified by a Markov chain
    transition matrix) to detect a possibly very weak signal. 
    The optimal detector is easily derived, and we focus on the harder problem of characterizing its performance through the
    (type-II) error exponent: the decay rate of the miss probability under a false alarm constraint. 
    The expression for the error exponent resembles the free energy of a spin glass in statistical physics, and we borrow
    techniques from that field to develop a lower bound. Our fully
    rigorous analysis uses large deviations theory to show that the lower bound exhibits a phase transition: strong
    performance is only guaranteed when the signal-to-noise ratio exceeds twice the entropy rate of the random walk.
    Monte Carlo simulations show that the lower bound fully captures the behavior of the true exponent.
\end{abstract}

\begin{IEEEkeywords}
Detecting random walks, combinatorial testing, error exponent, product of random matrices, Lyapunov exponent, random energy model, spin glasses, large deviations theory
\end{IEEEkeywords}

\ifCLASSOPTIONdraftcls
    \pagebreak[4]
\fi


\newcommand{\hadpow}[2]{#1^{(#2)}}
\newcommand{\pathrate}{\log \lambda_0}
\newcommand{\phiu}{\widetilde{\varphi}}
\newcommand{\st}{\vec{s}}   
\newcommand{\iid}{\overset{\text{i.i.d.}}{\sim}}
\newcommand{\indep}{\overset{\text{indep.}}{\sim}}

\section{Introduction}
Suppose we wish to make sense of a sequence of observations from nodes in a graph. The observations form a
spatiotemporal matrix, where each column contains the measurements at all nodes at a particular snapshot in time. As
illustrated in \fref{H0H1illus}, we need to distinguish between two hypotheses:
(a) every observation is just meaningless zero-mean Gaussian noise, or (b) an agent is undergoing a random walk on the
graph and the measurement at its location at each time has an elevated mean. We do not know the exact path of the agent, but we do
know its dynamics: with the graph structure assumed known, the agent's movements follow a well-defined finite-state Markov chain. In effect, we would like to exploit our knowledge of the graph structure (or the Markov chain) to help detect a possibly very weak signal.

In practice, this problem can arise from the detection of an intruder via a sensor network; the motion of a potential intruder might
be modeled as a random walk on a graph representing the network, and one is tasked with testing the hypothesis
that an intruder is currently present based on noisy measurements from each sensor. This kind of model has also been used in  
the detection of frequency-hopping or other highly oscillatory signals \cite{candes_detecting_2008}. More generally, it can be interpreted as the detection of a hidden Markov process, a problem with many applications (see, \emph{e.g.}, \cite{ting_near-optimal_2006,sung_neyman-pearson_2006,runkle_multi-aspect_2001,leong_error_2007,missaoui_land-mine_2011}.)

The task we have is a kind of \emph{combinatorial testing} problem 
\cite{addario-berry_combinatorial_2010, arias-castro_searching_2008,donoho_higher_2004,ingster_detection_2005}, in that there is an exponentially
large number of paths that could be anomalous. Thus, the alternative hypothesis is in fact a composite of
an exponentially large number of simple hypotheses. Despite this complexity, the optimal Neyman-Pearson detector in our problem turns out to be easy to derive and computationally tractable. However,
its performance is not so simple to characterize. 

We will use the (type-II) error exponent, which measures the rate of decay
of the miss detection probability when the false alarm probability is held fixed, as the performance metric.
One should expect it to depend on the signal-to-noise ratio (SNR) \emph{and} the degree to which the Markov
dynamics restrict the paths of the agent. If the SNR is too low, the true path will not be very different from
the noise. But if the number of potential paths is very small, it may be easy to rule out false alarms, and performance will be
better than when the number is very high. As the main focus of this paper, we will characterize the error exponent of the optimal detector and
quantify the above intuition. We do this by deriving a fully rigorous lower bound to the error exponent, using ideas borrowed from statistical
physics \cite{Mezard:1986fr, nishimori_statistical_2001, mezard_information_2009, merhav_statistical_2010}.

\begin{figure}
    \centering
    \begin{minipage}{0.8\textwidth}
    $\calH_0$: \\
    
    \vspace{-2ex}

    \includegraphics[width=5in]{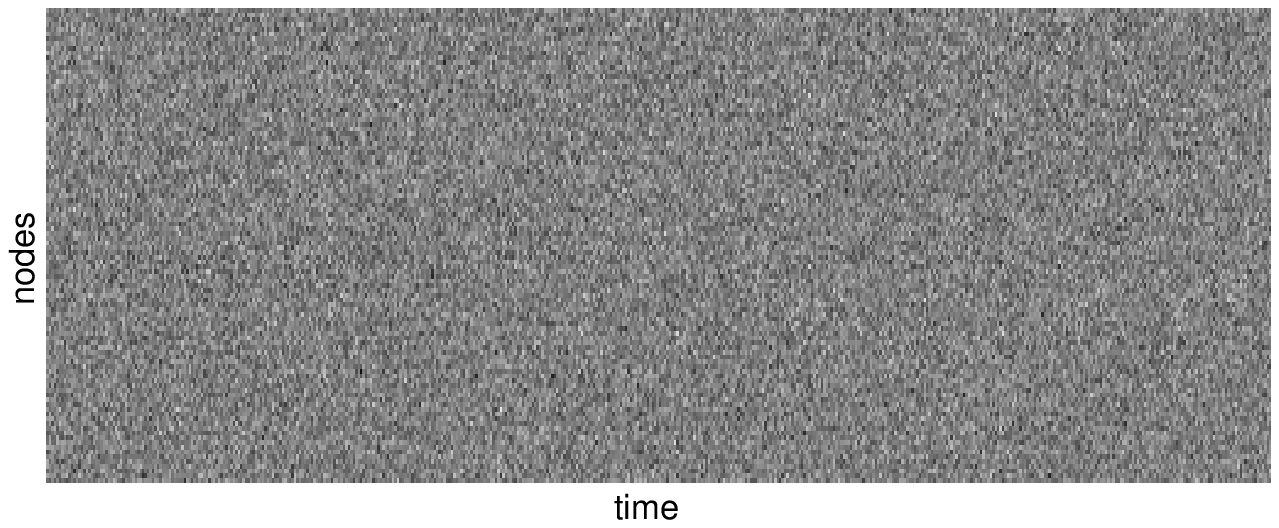} \\
    
    \vspace{-3ex}
    
    $\calH_1$: \\

    \vspace{-2ex}

    \includegraphics[width=5in]{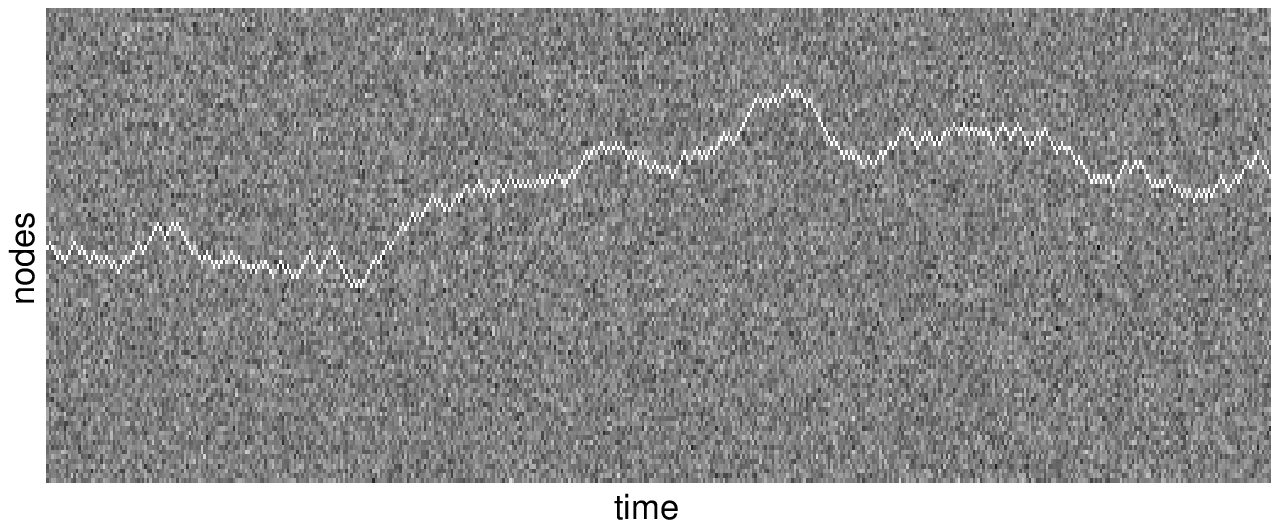}
\end{minipage}
\caption{ Illustration of the two hypotheses under consideration. Each column of the observation matrix shows the measurements at all nodes at a particular point in time.
              The null hypothesis $\calH_0$ (top) is that all of the measurements are just noise. The alternate hypothesis $\calH_1$ (bottom) is that a
          single node has an elevated mean at each time, and that node is chosen by a random walk. Here, we have illustrated a random walk on a line graph,
          but in this paper we consider the general case of any finite-state Markov chain.}
    \label{fig:H0H1illus}
\end{figure}

\subsection{Related and Prior Work}
Detecting a continuous Gauss-Markov process in Gaussian noise is a classical signal processing problem that
has been extensively studied (see, \emph{e.g.}, \cite{kailath_innovations_1970,scharf_likelihood_1977}.)
Hypothesis testing that tries to distinguish between two different finite-state Markov chains based 
on \emph{noiseless} realizations is also well-understood \cite{kazakos_bhattacharyya_1978,nakagawa_converse_1993,Levy:2008lr}.
In this work, we focus on the related problem of detecting random walks on directed and weighted graphs (which are finite state Markov chains)
based on noisy observations that are perturbed by additive Gaussian noise. These observations neither satisfy the Markov property nor are jointly Gaussian, making the problem a more difficult one.

There is some prior work on detecting hidden Markov processes such as the one we consider in this paper.
The structure of the optimal detector for a finite-state Markov chain in noise was addressed
in \cite{ting_near-optimal_2006,ting_detection_2006}. We are interested in going further and characterizing the asymptotic performance
of the optimal detector by computing the error exponent. For the Gauss-Markov case,
a closed-form expression for the error exponent was derived by Sung \textit{et al.} \cite{sung_neyman-pearson_2006} using a state space representation.
Our problem turns out to be more challenging. The error exponent, we shall see, is equal to the top Lyapunov exponent of the product of a sequence of random matrices \cite{furstenberg_products_1960, crisanti_products_1993}, a problem known to be difficult \cite{Tsitsiklis:1997qy}. Leong \textit{et al.} \cite{leong_error_2007} described a numerical technique to 
approximately compute the error exponent for detecting a two-state Markov chain in noise by discretizing a certain
integral equation. Unfortunately, numerical solutions based on discretization become computationally intractable for
general Markov chains with a large number of states, the case we address in this paper. In principle, one can always use
Monte Carlo simulations to estimate the Lyapunov exponent (and thus the error exponent.) However, they will not easily provide insights relating the error exponents to the SNR and the Markov chain structures.

Finally, we note that our problem is closely related to the general task of detecting nonzero-mean components of a Gaussian random vector 
\cite{donoho_higher_2004,ingster_detection_2005,arias-castro_near-optimal_2005}.
Addario-Berry \textit{et al.} characterized the performance in a very general setting \cite{addario-berry_combinatorial_2010}, 
bounding the Bayesian risk of the test; but in that work all of the nonzero-mean support sets under test are equiprobable and there is no Markov structure. 
Arias-Castro \textit{et al.} considered a problem similar to ours where a path on a graph has elevated mean while all other nodes are
zero-mean Gaussians \cite{arias-castro_searching_2008}; instead of a time series, they considered a single snapshot in
the asymptotic regime of very large graphs.

In this paper, we consider general graphs (or Markov chains) with an arbitrary number of nodes.
Drawing upon techniques originally developed in statistical physics
\cite{Mezard:1986fr, nishimori_statistical_2001, mezard_information_2009, merhav_statistical_2010}, we compute a lower bound on the
error exponent that appears in practice to be quite sharp. The lower bound exhibits a
\emph{phase transition} at a certain threshold SNR, separating the detectable and undetectable regimes. Some of
these results were previously presented in \cite{agaskar_detecting_2013,agaskar_optimal_2014}, but
we only justified them through nonrigorous arguments common in the statistical physics literature.
In this paper we use large deviations theory \cite{ellis_entropy_2005,dembo_large_2009} to provide a fully rigorous
derivation for the lower bound.

\subsection{Contributions}
We will precisely formulate the hypothesis testing problem in Section \ref{sec:math}, and introduce and motivate the error exponent
as the performance metric. The main contributions of the paper will follow: 

\begin{enumerate}[label=(\arabic*),wide=0pt]
    \item In Section \ref{sec:errexp}, 
we prove that the error exponent for this problem is well-defined and equal to the asymptotic Kullback-Leibler (KL) divergence rate of the two hypotheses.
We do this by generalizing the standard Chernoff-Stein lemma \cite{dembo_large_2009}, which gives the error exponent for independent
and identically distributed (i.i.d.) hypotheses, to the Markovian case.

\item Later in Section \ref{sec:errexp}, we develop upper and lower bounds for the error exponent. The upper bound is a simple genie bound. The lower bound is derived
borrowing techniques from statistical physics---it is related to the \emph{free energy density} of a new ``spin glass''
model \cite{Mezard:1986fr, nishimori_statistical_2001, mezard_information_2009, talagrand_mean_2010, merhav_statistical_2010}. 

\item We show how to explicitly compute the statistical physics-based lower bound. A rigorous proof of the expression is technical, so we present our results in two steps: first, we provide in Section~\ref{sec:nonrigorous} a high-level overview of our approach, emphasizing ideas and intuitions rather than rigor. 
Our discussions there also serve as a roadmap to the various results in Section \ref{sec:rigorous}, where we use large deviations theory to rigorously derive an expression for the lower bound and show how to compute it parametrically. The lower bound we derive exhibits a phase transition at an SNR equal to twice the entropy rate of the Markov chain. Below the threshold SNR, the bound is exactly equal to zero, indicating poor performance; above the threshold, there is rapid improvement in performance as the SNR increases.
        
\item In Section \ref{sec:simulations}, we compare the true error exponent (as estimated via Monte Carlo simulations) to the lower bound 
    and find that the bound fully captures its behavior, which appears to undergo a smoothed version of the phase transition at the predicted threshold. In the detectable SNR regime (above
    the threshold), our bound is also far better than an alternative bound obtained by ignoring the Markov structure, especially when the graph size is large.
\end{enumerate}

We offer some concluding remarks in Section \ref{sec:conclusions}.

\section{Problem Formulation}
\label{sec:math}

We consider testing the two hypotheses illustrated in \fref{H0H1illus}. The data form a matrix $\mY^N = [y_{m,n}]$ with $1 \leq m \leq M$
and $1 \leq n \leq N$, where $M$ is the number of nodes in the graph and $N$ is the number of observation times. As we allow the graph to be directed and weighted, the dynamics of an agent following a random walk on the graph can model any finite-state Markov chain. The two hypotheses are as follows:
\begin{align*}
     &\mathcal{H}_0:  y_{m,n} \iid \mathcal{N}(0, 1) \\
     &\mathcal{H}_1: \st = (s_1, s_2, \ldots, s_N) \sim \operatorname{Markov}(\mP) \\
     & \qquad y_{m,n} \vert \st \indep
     \begin{cases} 
         \mathcal{N}(\beta, 1), &\text{ if } m = s_n \\
         \mathcal{N}(0, 1), &\text{ if } m \neq s_n,
     \end{cases}
\end{align*}
where $\mP$ is the \emph{known} transition matrix of an irreducible and aperiodic $M$-state Markov chain [so that $\operatorname{Pr}(s_{n+1} = j  | s_n = i) = p_{i,j}$, the $ij$th entry of $\mP$]. 

Under the null hypothesis $\mathcal{H}_0$, the measurements are just i.i.d.~zero-mean standard Gaussian noise.
Under the alternate hypothesis $\mathcal{H}_1$, there is a sequence of states $\st = (s_1, s_2, \ldots, s_N) \in \{1,
\ldots, M\}^N$ produced by a Markov chain with transition matrix $\mP$, and we assume that $s_1$ is drawn from its
unique stationary distribution $\vec{\pi}$. By the Perron-Frobenius theorem for irreducible matrices
\cite{horn_matrix_2012}, the elements of $\vec{\pi}$ are all positive, meaning each state has a positive probability of
being initially chosen. Given the state sequence $\st$, the entries of the data matrix $\mY^N$ are still independent
Gaussian random variables. The difference is just that, in each column $n$ the Gaussian random variable at the $s_n$th
entry has an elevated mean $\beta$. This can be interpreted as the ``signature'' or ``evidence'' left behind by the agent. The variance in both hypotheses is set to $1$ without loss of generality; what matters is the signal to noise ratio (SNR) of $\beta^2$. In what follows, we will use $P_0(\cdot)$ and $P_1(\cdot)$ to refer to the probability
laws under $\calH_0$ and $\calH_1$, respectively, and $\EE_0$ and $\EE_1$ to refer to the corresponding expectation operators.

The optimal detector, that which minimizes the miss detection probability for a fixed
false alarm probability, is the Neyman-Pearson detector \cite{cover_elements_2006}.
The corresponding decision rule compares the likelihood ratio $L(\mY^N) \bydef \frac{P_1(\mY^N)}{P_0(\mY^N)}$ 
to a threshold and chooses $H_1$ only if it exceeds the threshold. The likelihood ratio for this problem can be computed as
\begin{align}
    L(\mY^N) & = \sum_{\st} P(\st) \frac{P_1(\mY^N | \st)}{P_0(\mY^N)} \nonumber \\
           & = \sum_{\st} P(\st) \exp\Big(\beta\sum_{n=1}^N y_{s_n,n} - \frac{N \beta^2}{2}\Big),\label{eq:L_sum}
\end{align}
where $P(\st) = \pi_{s_1} p_{s_1,s_2} \cdots p_{s_{N-1},s_N}$ is the probability of the state sequence $\st$ under the Markov chain $\mP$.
Conditioned on the state sequence $\st$, the variable $y_{m,n}$'s distribution is different under the two hypotheses only if $m = s_n$.
The expression in \eref{L_sum} might appear complicated, as the sum is over an exponentially large ($M^N$) number of
possible state sequences. However, the likelihood ratio turns out to be easy to compute: it was shown in
\cite{ting_near-optimal_2006} that $L(\mY^N)$ can be reformulated in terms of matrix products%
\footnote{Readers with a background in statistical physics may recognize this formula as an immediate consequence of the ``transfer matrix'' method \cite{Baxter:82}
as applied to a one-dimensional generalized Potts model with a quenched random field.}%
:
\begin{align}
    L(\mY^N) = \vec{\pi}^T \mD_1 \mP \mD_2 \mP \ldots \mP \mD_N \vec{1}, \label{eq:matrix_product_likelihood}
\end{align}
where $\mP$ is the transition matrix of the Markov chain, and $\mD_n$ is a diagonal matrix defined as
\begin{equation*}
\mD_n \bydef \exp\left(-\tfrac{\beta^2}{2}\right) \diag\Big(\exp(\beta y_{1,n}),\ldots, \exp(\beta y_{M,n})\Big)
\end{equation*}
for $1 \le n \le N$. Thus, the likelihood ratio can be computed in $\mathcal{O}(M^2 N)$ time.

A far more difficult problem is to characterize the performance of the detector, \emph{i.e.}, to compute the type-I (false alarm) error probability $\PF$ and the type-II (miss) error probability $\PM$. Under the optimal detector, these are given by the expressions
\begin{align}
        \PF & = \iint \cdots \int_{L(\mY^N) > \tau} P_0( \mY^N)\, d^{MN}\!\vy  \nonumber\\
        \PM & = \iint \cdots \int_{L(\mY^N) < \tau} P_1( \mY^N)\, d^{MN}\!\vy \nonumber
\end{align}
where $\tau$ is the Neyman-Pearson threshold chosen to achieve the constraint on $\PF$, and the integrals are over
all $MN$ variables $\{y_{m,n}\}$.
These are very high dimensional integrals for which only Monte Carlo techniques would be practical. However,
we would like to say something about the performance of these systems without having to simulate them. In particular,
we expect that the performance depends on two parameters: the element-wise SNR $\beta^2$, and some measure of
the complexity of the Markov chain $\mP$. For example, more restrictive dynamics for the state sequence $\st$ should make it easier
to correctly distinguish between the two hypotheses.

We consider the asymptotic performance of a detector as $N \to \infty$, \emph{i.e.}, as the observation time increases without bound. Let $\epsilon \in (0, 1)$ be a constant. Given a sequence of
optimal detectors $\delta_N(\mY^N)$ with false alarm constraint $P_{\text{false\_alarm}} \leq \epsilon$ (where $\delta_N$ has access to $N$ observations of the network),
the (type-II) \textit{error exponent} is
\begin{equation}\label{eq:eta}
    \eta \bydef -\lim_{N \to \infty} \frac{1}{N} \log P_{\text{miss}}(\delta_N).
\end{equation}
This means that $P_{\text{miss}}(\delta_N) = \exp(-\eta N + o(N))$, so that the dominant feature of the miss probability is that it decays exponentially with a rate of $\eta$. In the remainder of this paper, we will first prove that the error exponent in \eref{eta} is indeed a well-defined quantity, and then
explore techniques to analytically characterize it.

\section{The error exponent}
\label{sec:errexp}
\subsection{Existence}

The first question is whether the error exponent $\eta$ is a well-defined quantity. If $\calH_0$ and $\calH_1$ were both i.i.d.~hypotheses with single-letter marginal densities $p_0(\cdot)$ and $p_1(\cdot)$, then
the Chernoff-Stein lemma \cite{dembo_large_2009} would tell us that $\eta = D(p_0 || p_1) = -\EE_0 \log \frac{p_1(y)}{p_0(y)}$,
the Kullback-Leibler divergence of $p_1$ from $p_0$. However, since $\calH_1$ for our problem is not an i.i.d. hypothesis, the lemma in its original form is not applicable. So we prove the following generalization.
\begin{lemma}[Generalized Chernoff-Stein Lemma]
    \label{lemma:error_exponent_KL_rate}
    Suppose we have a sequence of hypotheses $\calH_0^N$ and $\calH_1^N$ with a well-defined Kullback-Leibler divergence rate
    \[
        \kappa \bydef -\lim_{N \to \infty} \frac{1}{N} \EE_0 \log\frac{P_1(\mY^N)}{P_0(\mY^N)} = -\lim_{N \to \infty} \frac{1}{N} \EE_0 \log L(\mY^N).
    \]
    Suppose furthermore that under
    $\calH_0$, the normalized log likelihood ratio $\ell_N \bydef \frac{1}{N} \log L(\mY^N)$ converges in probability to
    the limit of its expectation, $-\kappa$. Then the error exponent $\eta$ is well defined and $\eta = \kappa$.
\end{lemma}
\begin{IEEEproof}
    See Appendix \ref{ap:error_exponent_KL_rate}.
\end{IEEEproof}

To apply Lemma~\ref{lemma:error_exponent_KL_rate} to our problem, we need to verify that its assumptions hold. This is established by the following proposition, which uses results from the theory of matrix-valued stochastic processes \cite{furstenberg_products_1960}:
\begin{proposition}
    \label{prop:loglike_converges}
    The Kullback-Leibler divergence rate for our problem, 
    \[
        \kappa = -\lim_{N \to \infty} \frac{1}{N} \E_0 \log \left( \vec{\pi}^T \mD_1 \mP \mD_2 \mP \ldots \mP \mD_N \vec{1}\right),
    \]
    exists. Further, under $\calH_0$, the normalized log likelihood ratio converges almost surely: 
    \begin{equation}\label{eq:exp_as}
        \lim_{N \to \infty} \frac{1}{N} \log  \left( \vec{\pi}^T \mD_1 \mP \mD_2 \mP \ldots \mP \mD_N \vec{1}\right) \to -\kappa,
    \end{equation}
    and thus it converges in probability.
\end{proposition}
\begin{IEEEproof}
    We first note that, since $\mP$ is a stochastic matrix, we have $\mP \vec{1} = \vec{1}$, so we can add an extra factor of $\mP$ 
    into the expression (\ref{eq:matrix_product_likelihood}) to obtain
    \begin{equation}
        L(\mY^N) = \vec{\pi}^T \mD_1 \mP \cdots \mP \mD_N \mP \vone.
    \end{equation}
    Under $\calH_0$, the factors $\set{\mD_n \mP}_{n\geq 1}$ form an i.i.d. sequence of random matrices, with randomness induced by the Gaussian
    variables in the definition of $\mD_n$. In a classical paper \cite{furstenberg_products_1960}, Furstenberg and Kesten showed that
    for an i.i.d.~sequence of random matrices $\mX_n$, if $\E \log^+ \norm{\mX_n}_{\infty}$ is finite%
    \footnote{Here, $\log^+(x) = \max\{0, \log(x)\}$, and the matrix $\infty$-norm is induced by the $\ell^{\infty}$
    norm and is given by $\norm{\mX}_\infty \bydef \max_i \sum_j \abs{X_{i,j}}$.},
    the limit $\lim_{N \to \infty} \frac{1}{N} \E \log \norm{\mX_1 \cdots \mX_N}_{\infty}$ exists
    and the random quantity $\frac{1}{N} \log \norm{\mX_1 \cdots \mX_N}_\infty$ converges almost surely to the same limit. This quantity is
    equivalent to what is known as the (top) \emph{Lyapunov exponent}---the exponential rate of growth or decay of a product of random matrices.
    First, let us show that the result applies to the factors $\set{\mD_n \mP}$. For any fixed $n$, we have:
    \begin{align*}
        \E \log^+ \norm{\mD_n \mP}_{\infty} & \leq \E \Big| \log \, \norm{\mD_n \mP}_{\infty} \Big| \\ 
        & = \E \Big| \log\max_m \Big\{\exp( \beta y_{m,n} - \frac{\beta^2}{2} )\Big\}   \Big| \\
        & = \E \Big| \beta \max_m y_{m,n} - \frac{\beta^2}{2}  \Big| < \infty.
    \end{align*}
So the condition we
    need to apply the Furstenberg-Kesten result holds.
    Now we must relate the likelihood ratio to the norm of the product of random matrices.
    Using H\"older's inequality, we have
    \begin{align}
       &  \vec{\pi}^T \mD_1 \mP \cdots \mP \mD_N \mP \vone \nonumber \\
        & \leq \norm{\vec{\pi}}_1 \norm{ \mD_1 \mP \cdots \mP \mD_N \mP \vone}_\infty \nonumber \\
        & = \norm{ \mD_1 \mP \cdots \mP \mD_N \mP }_\infty, \label{eq:likelihood_ub:2}
    \end{align}
    where (\ref{eq:likelihood_ub:2}) follows from the definition of the matrix $\infty$-norm and the fact that all of the matrices
    involved are nonnegative and all of the vectors are positive.
    Meanwhile, as a lower bound, we let $\pi_{\min} = \min_m \pi_m$ and it holds that
    \begin{align}
        & \vec{\pi}^T \mD_1 \mP \cdots \mP \mD_N \mP \vone \nonumber \\
        & \geq \pi_{\min} \norm{\mD_1 \mP \cdots \mP \mD_N \mP \vone}_1 \nonumber \\ 
        & \geq \pi_{\min} \norm{\mD_1 \mP \cdots \mP \mD_N \mP }_\infty \label{eq:likelihood_lb:3},
    \end{align}
    where \eref{likelihood_lb:3} again follows from the definition of the matrix $\infty$-norm. So we can sandwich the
    log likelihood ratio to within a vanishing constant:
    \begin{equation*}
        \frac{1}{N} \log \norm{ \mD_1 \mP \cdots \mD_N \mP}_\infty + \frac{1}{N} \log \pi_{\min}  
        \leq \frac{1}{N} \log L(\mY^N) \leq  
        \frac{1}{N} \log \norm{ \mD_1 \mP \cdots \mD_N \mP}_\infty.
    \end{equation*}
    The outer expressions converge almost surely and in expectation due to Furstenberg and Kesten's results 
    \cite[Theorems 1 and 2]{furstenberg_products_1960}, so the log likelihood ratio
    must converge in the same way.
\end{IEEEproof}

\begin{remark}
Note that the proof only requires that the probability distribution of the initial state $s_1$ be positive at all
nodes---there is no need to start from the stationary distribution $\vec{\pi}$. In fact, since $\mP$ is irreducible and aperiodic,
we could relax the positivity constraint on the initial distribution and start with any known distribution.
\end{remark}

Lemma \ref{lemma:error_exponent_KL_rate} and Proposition \ref{prop:loglike_converges}
indicate that computing the error exponent boils down to computing the top Lyapunov exponent of products of random matrices, a problem known to be hard \cite{Tsitsiklis:1997qy}. For $M \times M$ matrices, it generally requires solving an integral equation to obtain the invariant measure of a continuous diffusion 
process on a $M$-dimensional real projective space \cite{comtet_products_2010}. In low dimensions (\emph{e.g.}, $M = 2$ or $3$), this can be done with 
numerical quadrature (see, \emph{e.g.}, \cite{leong_error_2007, jacquet_entropy_2008}), but this becomes intractable for high dimensional
problems. Thanks to almost sure convergence of the normalized partial products in \eref{exp_as}, one can use Monte Carlo simulations to estimate the error exponents. A simple Monte Carlo procedure that does just that is presented in Section \ref{sec:simulations}, where we report some results of numerical simulations.

\subsection{Upper and Lower Bounds}
\label{subsec:bounds}

Obtaining analytical expressions for the error exponents for general Markov chain structures is expected to be a very challenging task. Instead, we will focus on deriving bounds for the error exponents. The Lyapunov exponent formulation of the error exponent as given in \eref{exp_as} does not lend itself to easy analysis. To proceed, we use the alternative form of the likelihood ratio in \eref{L_sum} to rewrite the error exponent as follows
\begin{align}
    \eta & = \lim_{N \to \infty} -\frac{1}{N} \EE \log \left( \sum_{\st} P(\st) \exp\left(\beta y_{\st} - \frac{N\beta^2}{2}\right) \right) \nonumber\\
    & = \frac{\beta^2}{2} - \underbrace{\lim_{N\to\infty}\frac{1}{N} \EE \log \Big( \sum_{\st} P(\st) \exp(\beta y_{\st} )   \Big)}_{\varphi(\beta)},\label{eq:eta_phi}
\end{align}
where $\st = (s_1, s_2, \ldots, s_N) \in \{1, \ldots, M\}^N$ is a state sequence of the Markov chain, and we define
\begin{equation}\label{eq:y_s}
y_{\st} \bydef \sum_{n=1}^N y_{s_n,n} \sim \mathcal{N}(0,N)
\end{equation}
to be the sum of the Gaussian random variables associated with a given state sequence $\st$.
Here, and in what follows, we shall simply use $\EE$ to refer to the expectation under $\calH_0$, since we have no further use for $\EE_1$.
To study the behavior of the error exponent, we just need to study
\begin{equation}\label{eq:phi}
    \varphi(\beta) = \lim_{N \to \infty} \frac{1}{N} \EE \log \Big(\sum_{\st} P(\st) \exp(\beta y_{\st}) \Big).
\end{equation}

We will derive upper and lower bounds on this quantity, which will translate into bounds on the error exponent $\eta$. There is a simple lower bound: by treating the sum $\sum_{\st} P(\st) \exp(\beta y_{\st})$ as an expectation and applying Jensen's inequality, we get
\begin{equation*}
    \varphi(\beta) \geq \lim_{N \to \infty} \frac{1}{N} \EE \sum_{\st} P(\st) \log \exp(\beta y_{\st}) = 0.
\end{equation*}
This then gives us an upper bound for the error exponent
\[
\eta \leq \frac{\beta^2}{2},
\]
which can also be interpreted as the ``genie'' bound: if we are given
the true state sequence $\vec{s}$, then we can examine only the variables along that path and ignore all others, leading to an i.i.d.~hypothesis
testing problem with error exponent $\frac{\beta^2}{2}$. It provides an upper bound on the true error exponent since the extra side information about the correct path can only improve the performance. 

To get a lower bound on $\eta$, we can still apply Jensen's inequality, but this time to the outer expectation $\EE$ in \eref{phi}, to obtain
\[
    \varphi(\beta) \leq \lim_{N \to \infty} \frac{1}{N} \log\Big( \sum_{\st} P(\st) \E \exp(\beta y_{\st})\Big) = \frac{\beta^2}{2},
\]
which gives us $\eta \geq 0$. Of course, this is trivial since $\eta$ is equal to a limit of Kullback-Leibler divergences, which are always nonnegative.
Another lower bound can be obtained by considering the test statistics $y_n = \sum_m y_{mn}$, the sums of the states in
each time step. Since we are discarding information, the error exponent for this problem can be no greater than that for the
original problem. But the new problem is just testing two i.i.d. hypotheses
$y_n \iid \NN(0, M)$ and $y_n \iid \NN(\beta, M)$. As we know, in the i.i.d.~case the error exponent is simply the Kullback-Leibler divergence of these
two densities, giving us a lower bound of
\[
    \eta \geq \frac{\beta^2}{2 M}.
\]
This is a nontrivial bound, but just barely. For large $M$, the error exponent is very small indeed. In fact, we would
need $M$ times the observation length to obtain the same performance as the genie-aided detector.

We will spend the remainder of this section and all of the next two sections computing a nontrivial lower bound for $\eta$,
one that we will find empirically to fully capture its behavior. Qualitatively, this lower bound will guarantee that,
above a certain threshold SNR, the error exponent will be bounded by
\[
    \eta \geq \frac{\beta^2}{2} - O(\beta),
\]
meaning to leading order, the maximum likelihood detector will be just as good as the genie-aided detector.

To develop this bound, we will borrow ideas from the theory of spin glasses
\cite{talagrand_mean_2010, mezard_information_2009, Mezard:1986fr, nishimori_statistical_2001,merhav_statistical_2010},
a class of disordered systems studied in statistical physics. In fact, we have already chosen our notation so that our
result closely resembles the quantities studied in that field.
In particular, the function $\varphi(\beta)$ resembles the so-called ``free energy density'' of a spin glass, defined as
\begin{equation}\label{eq:free_energy}
    \phi(\beta) = -\frac{1}{\beta} \lim_{N \to \infty} \frac{1}{N} \E \log \Big(\sum_{\st} \exp(-\beta H(\st)) \Big),  
\end{equation}
where $N$ is the number of particles in the spin glass, $\st \in \R^N$ is an indexing vector representing the configurations of the system (there are typically exponentially large number of them), $\beta$ is the inverse temperature parameter, and $H(\cdot)$ is a random Hamiltonian, a function defining the energy of each configuration. For our problem, we can write the function in \eref{phi} as $\varphi(\beta) = -\beta \phi(\beta)$ if we choose the Hamiltonian to be
\begin{equation}\label{eq:Hamiltonian}
H(\st) = -y_{\st} - \frac{1}{\beta} \log P(\st).
\end{equation} 
Despite the extra factor of $-\beta$, to be concise we will abuse the terminology and henceforth refer to $\varphi(\beta)$ for our problem as the
free energy density.

Computing the free energy density of a disordered system is often very difficult. In fact, there are seemingly simple models that have been studied
for many years with no exact solution \cite{talagrand_mean_2010, bolthausen_spin_2007}. The main challenge lies in the fact that the free energy density $\phi(\beta)$ in \eref{free_energy} involves the sum of an exponentially large number of random variables. The high-dimensional \emph{correlation structures} of the random Hamiltonians $\set{H(\st)}_{\st}$ can often lead to remarkable phenomena (see, \emph{e.g.}, \cite{mezard_information_2009, talagrand_mean_2010, panchenko_sherrington-kirkpatrick_2013}).

In our problem, the correlations of the Hamiltonians can be computed as follows. Let $\st^1, \st^2$ denote two arbitrary paths of the Markov chain, and let $H(\st^1), H(\st^2)$ be the associated Hamiltonians as defined in \eref{Hamiltonian}. Using \eref{y_s}, we can easily verify that
\begin{equation}
\operatorname{cov}(H(\st^1), H(\st^1)) = \EE \, y_{\st^1} y_{\st^2} = \sum_{n=1}^N \charfn(s_n^1 = s_n^2),
\label{eq:ys_covariance}
\end{equation}
where $\charfn(\cdot)$ is the indicator function. This means that the Hamiltonians of the various states in our problem are indeed correlated, and the covariance is equal to the number of times the two sequences overlap. 

In the spin glass literature, removing or just reducing the correlations between state Hamiltonians can often simplify a
problem \cite{derrida_random-energy_1981,talagrand_mean_2010}. We follow this idea: if we drop the correlations, we obtain a modified function%
\footnote{Strictly speaking, we need to show that $\phiu(\beta)$ exists, \emph{i.e.} that the limit is actually well-defined.
We will do this in Section \ref{sec:rigorous} by actually computing it. Until then, we presuppose its existence in all our arguments.}
\begin{equation}\label{eq:phiu}
    \phiu(\beta) = \lim_{N \to \infty} \frac{1}{N} \EE \log \Big( \sum_{\st} P(\st) \exp(\beta x_{\st})  \Big),
\end{equation}
where $x_{\st} \iid \NN(0, N)$, \emph{i.e.} they are an uncorrelated Gaussian ensemble with the same variance as the $y_{\st}$.
We note that the two functions in \eref{phiu} and \eref{phi} have exactly the same form, the only difference being the 
absence of correlation in $\set{x_{\st}}$. Dropping the correlation, as we shall see, makes our problem tractable%
\footnote{In spin glass parlance, our function $\phiu(\beta)$ may be regarded as the (rescaled) free energy density of a new generalization of the random energy model (REM) \cite{derrida_random-energy_1981,talagrand_mean_2010}.}%
.
Interestingly, it also provides a lower bound on the error exponent, which is precisely what we seek for our problem. The argument relies on the following lemma:

\begin{lemma}[{Slepian's Lemma \cite[pp. 12--15]{talagrand_mean_2010}}]
    Let the function $F : \R^L \to \R$ (for some $L$) satisfy the moderate growth condition
    \begin{equation*}
        \lim_{\norm{\vv} \to \infty} F(\vv) \exp(-a \norm{\vv}^2) = 0 \text{ for all } a > 0,
    \end{equation*}
    and have nonnegative mixed derivatives:
    \begin{equation*}
        \frac{\partial^2 F}{\partial v_i \partial v_j} \geq 0 \text{ for } i \neq j.
    \end{equation*}
    Suppose that we have two independent zero-mean Gaussian random vectors $\vx$ and $\vy$ taking values in
    $\R^L$ such that $\E x_i^2 = \E y_i^2$ and $\E y_i y_j \geq \E x_i x_j$ for $i \neq j$. Then $\E F(\vy) \geq \E
    F(\vx)$.
\end{lemma}

Applying this to $\varphi(\beta)$ gives us the desired lower bound on the error exponent:
\begin{proposition}
The error exponent satisfies $\eta \geq \frac{\beta^2}{2} - \phiu(\beta)$.
\end{proposition}
\begin{IEEEproof}
    Define $F(\vv) = -\log(\sum_{\st} P(\st) \exp(\beta v_{\st}))$. This is a function from $\R^{M^N}$ to $\R$ that
    clearly satisfies the moderate growth condition. We can compute the cross second derivative with respect to
    $v_{\st^1}$ and $v_{\st^2}$, with $\st^1 \neq \st^2$, as:
    \begin{equation*}
        \frac{\partial F}{\partial v_{\st^1} \partial v_{\st^2}} =
             \displaystyle\frac{ \beta^2 P(\st^1) P(\st^2) \exp(\beta(v_{\st^1} + v_{\st^2}))}{[\sum_{\st} P(\st) \exp(\beta v_{\st})]^2},
    \end{equation*}
    which is clearly nonnegative. From (\ref{eq:ys_covariance}), we know that for $\st^1 \neq
    \st^2$, $\E y_{\st^1} y_{\st^2} \geq 0$, and we have constructed the $\vx$ ensemble so that $\E x_{\st^1} x_{\st^2} = 0$.
    Thus, applying Slepian's Lemma gives us $\E F(\vy) \geq \E F(\vx)$, which is equivalent to $\varphi(\beta) \leq \phiu(\beta)$.
    The statement of the proposition then follows immediately from \eref{eta_phi}. 
\end{IEEEproof}

Next, we will show how to explicitly compute $\phiu(\beta)$ by using tools from large deviations theory. Before delving into the technical results, we first present in \sref{nonrigorous} a high-level and non-rigorous overview of the main ideas used in our approach. The discussions there also provide a roadmap to the various rigorous arguments that lead to our final results, stated as Theorem~\ref{theorem:rigorous_saddlept} and Propositions~\ref{prop:errexp_bound_regular} and \ref{prop:errexp_bound_general} in \sref{rigorous}.

\section{Main Ideas and Roadmap to the Technical Results}
\label{sec:nonrigorous}

To begin, we can rewrite the free energy density as:
\begin{equation}
    \phiu(\beta) = \lim_{N \to \infty} \frac{1}{N} \EE \log \sum_{\st \in \calP^N} \exp(\beta x_{\st} + \log P(\st)), \label{eq:phiu_logPS_in_exponent}
\end{equation}
where we are considering only the set $\calP^N \subset \{1,\ldots,M\}^N$ of paths that have nonzero 
probability under the Markov chain $\mP$ (the other paths contributed nothing to the sum in the first place.)

We can group the terms of the sum by their $\frac{1}{N}\log P(\st)$ and $\frac{1}{N} x_{\st}$ values, dividing them into bins with a small width $\delta$.
Counting the number of configurations (\emph{i.e.}, paths) in each bin as
\begin{equation*}
    C^{\delta}_N(\rho, \xi)  \bydef 
       \#\{ \st \in \calP^N: \log P(\st) \in [N \rho, N(\rho + \delta)]
            \text{ and } x_{\st} \in [N \xi, N(\xi + \delta)]  \},
\end{equation*}
then we should be able to approximate the sum as
\begin{equation}
    \phiu(\beta) \approx \lim_{N \to \infty} \frac{1}{N} \EE \log \sum_{\rho} \sum_{\xi} C^\delta_N(\rho, \xi) \exp(N[ \beta \xi + \rho] ),
    \label{eq:phiu_gridded}
\end{equation}
where the sums are over cornerpoints of the bins. In Section~\ref{sec:rigorous}, we will show that a form of this approximation can be made exact.
\begin{figure*}
    \centering
    \includegraphics{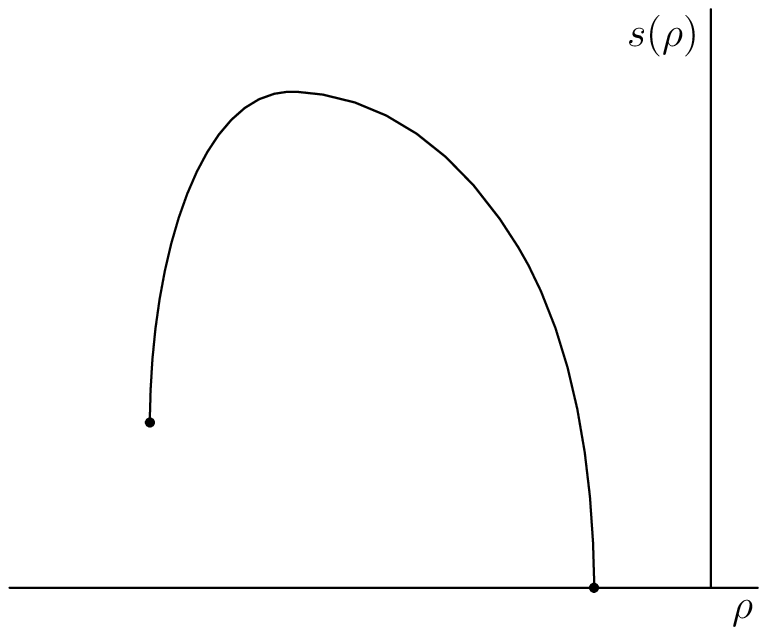}
    \hspace{.2in}
    \includegraphics{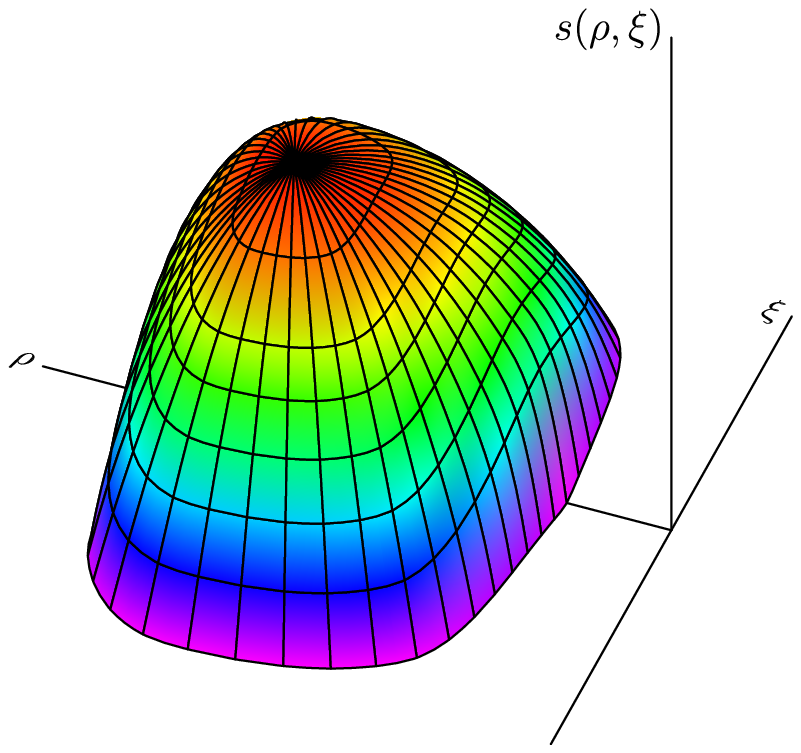}
    \caption{Notional illustrations of the microcanonical entropy densities $s(\rho)$ (left) and $s(\rho, \xi)$ (right). $s(\rho)$ is the exponential
             growth rate of the number of paths $\st$ satisfying $\frac{1}{N} \log P(\st) \approx \rho$, whereas $s(\rho, \xi)$ is, with probability 1, the exponential growth rate
             of the number of paths satisfying $\frac{1}{N} \log P(\st) \approx \rho$ and $\frac{1}{N} x_{\st} \approx \xi$. The density function $s(\rho, \xi)$ has a
compact support, outside of which the density $s(\rho, \xi) = -\infty$, meaning that there is no path there. Analytical expressions for these functions
             are derived in Section~\ref{sec:rigorous}.}
    \label{fig:notional_J_and_srx}
\end{figure*}

Of course, $C^\delta_N(\cdot, \cdot)$ is random due to its dependence on the Gaussian variables $\{x_{\st}\}$, but it
turns out that there will be a concentration of measure phenomenon that will allow us to treat it deterministically in the large $N$ limit. 
If we consider only the marginal count $C_N^\delta(\rho)$ of paths satisfying $\log P(\st) \in [N \rho, N (\rho + \delta)]$,
then there is no randomness involved; we can show that this count grows exponentially:
\[
C_N^\delta(\rho) = \exp\Big(N[ \sup_{\rho' \in [\rho, \rho+\delta]}s(\rho')] + o(N)\Big),
\]
where $s(\rho)$ is the ``microcanonical entropy density'' function for $\frac{1}{N} \log P(\st)$.
This is physics jargon for the exponential growth rate of the number of configurations within an energy level \cite{mezard_information_2009}.
In Section \ref{sec:rigorous}, we will show how to compute it (see Proposition \ref{prop:LDP_rho}) and derive several important properties (see Proposition \ref{prop:Jprops}).
A notional illustration based on those properties is provided in \fref{notional_J_and_srx}.

Meanwhile, the full count $C^\delta_N(\cdot, \cdot)$ will also grow exponentially:
\[
    C^\delta_N(\rho, \xi) = \exp\Bigg(N \Bigg[\sup_{\substack{\rho' \in [\rho, \rho + \delta]\\ \xi' \in [\xi, \xi+\delta] }} s(\rho',\xi')\Bigg] + o(N)  \Bigg)
\]
with probability $1$ under the distribution of the $x_{\st}$, where $s(\rho,\xi)$ is the two-dimensional microcanonical entropy density function for the pair $(\frac{1}{N} \log P(\st), \frac{1}{N}x_{\st})$.
In Section \ref{sec:rigorous}, we will show how to compute $s(\rho, \xi)$ (see Theorem \ref{theorem:LDP_rhoxi}), which is of course closely related to $s(\rho)$. 
Again a notional illustration is provided in Figure \ref{fig:notional_J_and_srx}.

As $N$ grows, the number of states grows exponentially, and we can let the bin width $\delta$ vanish and approximate the sum (\ref{eq:phiu_gridded}) by an integral.
The free energy density can then be evaluated as
\begin{align}
\phiu(\beta)  & \approx  \lim_{N \to \infty} \frac{1}{N} \EE \log \iint  \exp\left(N[s(\rho,\xi) + \beta \xi + \rho] \right)  d\rho d\xi \nonumber\\
& = \sup_{\rho, \xi}  \Big\{s(\rho,\xi) + \beta \xi + \rho\Big\},\label{eq:Laplace}
\end{align}
where the equality is obtained via the \emph{Laplace principle}%
\footnote{The Laplace principle states that when $N$ is very large, $\int \exp(N f(x)) dx = \exp(N \sup_x f(x) + o(N))$,
    \emph{i.e.} the integral is dominated by the peak. This is also known as the \emph{saddle-point technique},
    a powerful tool in asymptotic integration \cite{bruijn_asymptotic_2010}.}
\cite{bruijn_asymptotic_2010}; we will use a rigorous formulation of this principle in Theorem \ref{theorem:rigorous_saddlept} in the next section. 

\begin{figure}[t]
   \centering
    \includegraphics{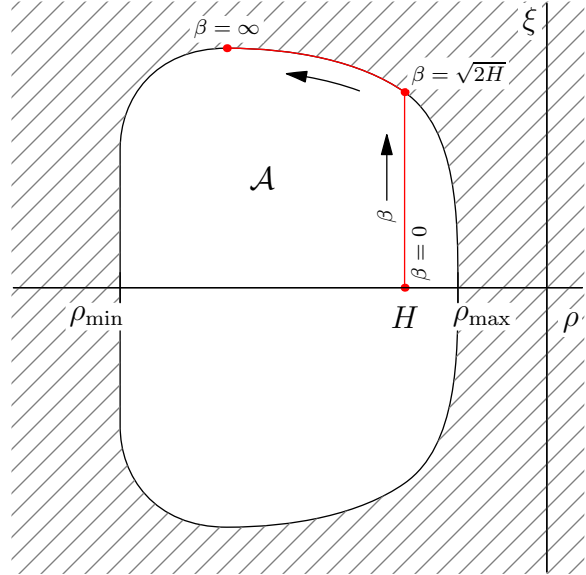}
    \caption{The location of the supremum that ultimately gives us $\phiu(\beta)$ is illustrated here. The entropy density $s(\rho, \xi)$ is finite only in the compact region
        $\calA$ illustrated here---this is also the effective domain of the large deviation rate function $I(\rho, \xi)$, which will be defined in (\ref{eq:Irhoxi_def}).
             Below the threshold, the supremum in \eref{Laplace} is achieved at a critical point in the interior; above
             the threshold, the supremum moves along the boundary as $\beta$ increases. The change in behavior at the
             threshold leads to a phase transition. Technical details will be provided in Section \ref{sec:rigorous}.}
     \label{fig:s_argmax}
\end{figure}

To actually compute $\phiu(\beta)$, we will need to evaluate the supremum in \eref{Laplace}. As it turns out, the microcanonical entropy density $s(\rho, \xi)$ has a compact support (see Figures~\ref{fig:notional_J_and_srx} and \ref{fig:s_argmax}), outside of which the density $s(\rho, \xi) = -\infty$. The supremum can thus be only achieved at the interior or the boundary of the support region. As illustrated in Figure~\ref{fig:s_argmax}, the location where the supremum is achieved depends on whether $\beta$ is greater or less than a threshold of $\sqrt{2 H}$, where $H$ is the entropy rate of the Markov chain $\mP$ (defined in Section \ref{sec:rigorous}.) As shown in the figure, below the threshold, the supremum is achieved at a critical point in the interior of the support region; as $\beta$ increases the critical point moves up along the line $\rho = H$ until it hits the boundary. As $\beta$ continues to increase beyond the threshold, the location of the supremum moves along the boundary in a direction of decreasing $\rho$.  The change in behavior at the threshold corresponds to a \emph{phase transition} in $\phiu(\beta)$. In Section~\ref{subsec:evaluating_phiu} we will provide a closed-form expression for $\phiu(\beta)$ below the threshold, and a parametric representation for it above the threshold. The reader who wishes to skip the technical details can skip directly to that section, where we provide these expressions.

\section{Rigorous Derivation}
\label{sec:rigorous}

In this section, we use results from large deviations theory to rigorously derive expressions for the lower bound.

\subsection{Large deviations and the microcanonical entropy density}
First, we introduce the large deviations property for a sequence of probability measures:
\begin{definition}[Large Deviation Property {\cite[pp. 35-36]{ellis_entropy_2005}}]
Let $\calX$ be a complete separable metric space and  $\calB(\calX)$ be the Borel $\sigma$-field of $\calX$,
Then the sequence $\{Q_N\}_{N=1}^\infty$ of probability measures on $\calB(\calX)$ satisfies
the \emph{large deviations property} if there is a lower semicontinuous function $I : \calX \to [0, \infty]$ (the function may take
the value $\infty$) with compact level sets such that
\begin{enumerate}
    \item $\displaystyle\limsup_{N \to \infty} \frac{1}{N} \log Q_N(B) \leq -\inf_{\vx \in B} I(\vx)$ for every closed set $B$ in $\calB$, and
    \item $\displaystyle\liminf_{N \to \infty} \frac{1}{N} \log Q_N(U) \geq -\inf_{\vx \in U} I(\vx)$ for every open set $U$ in $\calB$.
\end{enumerate}
$I(\vx)$ is  known as the \emph{rate function}. 
\label{def:ldp}
\end{definition}

To apply large deviations theory, we will consider the ordered pairs $(\frac{1}{N} \log P(\st), \frac{1}{N} x_{\st})$ for $\st \in \calP^N$
as inducing an empirical measure $Q_N$; for any set $B \subset \R^2$,
\[
Q_N(B) \bydef \frac{1}{\#\calP^N} \#\left\{\st\in\calP^N : \left(\frac{1}{N} \log P(\st), \frac{1}{N} x_{\st}\right) \in B \right\}. 
\]
One way to think about this is as follows: if we choose an allowable state $\st \in \calP^N$ uniformly at random
(rather than choosing it by running the Markov chain), then $Q_N(B)$ is the probability that the ordered pair
$(\frac{1}{N} P(\st), \frac{1}{N} x_{\st})$ is in $B$. 
This is just the number of states in the set $B$ divided by the total number of allowable paths $\# \calP^N$.  

Since the $\{x_{\st}\}$ are random, $Q_N$ itself is a \emph{random} probability measure. It is important
to note that there are two levels of randomness here: first, the random variables $\{\frac{1}{N} x_{\st}\}$ themselves, and 
second, the empirical probability distribution $Q_N$ that they induce when paired with the log probabilities $\frac{1}{N} \log P(\st)$.
We will show that \emph{with probability 1}, the empirical probability measure will satisfy the large deviations property
in Definition \ref{def:ldp}, and we will compute the rate function $I(\rho, \xi)$.

We will need to compute $\# \calP^N$, the number of allowable paths.
If every entry of transition matrix $\mP$ is nonzero, then this is simple: $\#\calP^N = M^N$.
If each row of $\mP$ has exactly $K$ nonzero entries, meaning that each state can transition
to only $K$ other states, then $\#\calP^N = K^N$. However, in the general case, we have:
\[
\begin{aligned}
    \#\calP^N & = \sum_{s_1} \sum_{s_2} \cdots \sum_{s_N} \charfn(p_{s_1,s_2} \neq 0) \charfn(p_{s_2,s_3} \neq 0) \cdots \charfn(p_{s_{N-1},s_N} \neq 0) \\
    & = \vec{1}^T\left(\hadpow{\mP}{0}\right)^{N-1} \vec{1},
\end{aligned}
\]
where for any matrix $\mA$ and $t\in \R$, we define $\hadpow{\mA}{t}$ to be the sparsity-preserving \emph{Hadamard power} of $\mA$, whose $ij$th entry is
given by:
\[
    [\hadpow{\mA}{t}]_{i,j} = \begin{cases}
        [\mA]_{i,j}^t & \text{if } [\mA]_{i,j} \neq 0 \\
        0         & \text{if } [\mA]_{i,j} = 0.
    \end{cases}
\]
In particular, $\hadpow{\mP}{0}$ is a $0$-$1$ matrix that is the adjacency matrix of the directed graph underlying the Markov chain. Its $ij$th element is 1
if and only if there is a nonzero probability of transitioning to state $j$ directly from state $i$. Since $\mP$ is
irreducible and aperiodic, so must be $\hadpow{\mP}{0}$. Due to the Perron-Frobenius theorem, $\lmax(\hadpow{\mP}{0})$
is simple, the associated left and right eigenvectors can be chosen to be positive, and all other eigenvalues are of smaller
magnitude, so we can see that $\#\calP^N$ grows exponentially with rate
\[
    \lim_{N \to \infty} \frac{1}{N} \log \#\calP^N = \log \lmax(\hadpow{\mP}{0}).
\]

The first step toward showing that $Q_N$ satisfies the large deviation property with probability 1 is to show that 
its marginal $Q^1_N$ with respect to the first argument satisfies the large deviation property.
This is simply the empirical probability measure on $\R$ induced by
$\frac{1}{N} \log P(\st)$ for all $\st \in \calP^N$.
It is \emph{not} a random measure, since it does not depend on the Gaussian random variables $\{x_{\st}\}$.
We will exploit the powerful G\"artner-Ellis theorem:
\begin{theorem}[{G\"artner-Ellis Theorem \cite[p. 47]{ellis_entropy_2005}}]
    Suppose we have a sequence of random variables $X_N$ taking values in $\R$. Let $\frac{1}{N}\log \E \exp(t X_N)$ be 
    finite for every $t, N$. Suppose the limiting cumulant generating function (CGF), given by $c(t) \bydef \lim_{N \to \infty}
    \frac{1}{N} \log \E \exp(t X_N)$, exists and is finite and differentiable for all $t$.
    Then the Legendre-Fenchel transform of $c(t)$, given by
    \[
        I(x) = \sup_{t \in \R} \, \Big\{tx - c(t)\Big\},
    \]
    is convex, lower semicontinuous, nonnegative, has compact level sets, satisfies $\inf_x I(x)$ = 0, and is the large deviations rate function
    for $\frac{1}{N} X_N$.
\end{theorem}

In our case, the random variable $X_N$ is the one induced by choosing a state $\st$ uniformly at random from $\calP^N$, and
taking $X_N = \log P(\st)$. We can compute the limiting CGF as:
\begin{align}
    c(t) & = \lim_{N \to \infty}\frac{1}{N} \log \Bigg( \frac{1}{\vec{1}^T\left(\hadpow{\mP}{0}\right)^{N-1} \vec{1}} \sum_{s_1}\cdots\sum_{s_N} \pi_{s_1}^t 
     p_{s_1,s_2}^t\charfn(p_{s_1,s_2} \neq 0) \cdots p^t_{s_{N-1},s_N}\charfn(p_{s_{N-1},s_N} \neq 0) \Bigg) \nonumber \\ 
    & = \lim_{N \to \infty} \frac{1}{N} \log \left[\left(\hadpow{\vec{\pi}}{t}\right)^T \left( \hadpow{\mP}{t} \right)^{N-1} \vec{1} \right]
        - \lim_{N \to \infty}\frac{1}{N} \log \left[\vec{1}^T \left( \hadpow{\mP}{0} \right)^{N-1} \vec{1} \right] \nonumber \\
    & = \log \lmax(\hadpow{\mP}{t}) - \log \lmax(\hadpow{\mP}{0}),
    \label{eq:logP_cgf}
\end{align}
again using the Perron-Frobenius theorem, which due to the irreducibility and aperiodicity of $\mP$ ensures that only the top eigenvalue remains for both terms.
To apply the G\"artner-Ellis theorem, we need to show that $c(t)$ is differentiable. This follows
from the following proposition, which provides several properties of 
the function $\log \lmax(\hadpow{\mP}{t})$ that we will need. To simplify the notation, we will define 
\[
\lambda_t \bydef \lmax(\hadpow{\mP}{t}).
\]
\begin{proposition}
    The function $\log \lambda_t$ satisfies the following properties:
    \begin{enumerate}[label=(\arabic*),wide=0pt]
        \item $\log \lambda_t$ is finite, analytic, and convex on $\R$.
        \item $\log \lambda_t$ is in fact \emph{strictly} convex on $\R$ unless 
            $\mP$ is the transition matrix for a uniform random walk on a regular graph, \emph{i.e.} there is some
            integer $K \leq M$ such that each row of $\mP$ has exactly $K$ nonzero entries, all of which are
            $\frac{1}{K}$. In that case, $\log \lambda_t = (1-t) \log K$.
        \item Let $\va_t$ and $\vb_t$ be the left and right Perron-Frobenius eigenvectors of $\hadpow{\mP}{t}$, respectively.
              Then the derivative is given by:
              \begin{equation}
                  \frac{d}{dt}\log \lambda_t = \frac{\va_t^T [(\log \mP) \circ \hadpow{\mP}{t}] \vb_t}{\va_t^T [\hadpow{\mP}{t}] \vb_t}, \label{eq:loglt_deriv}
              \end{equation}
            where the $\log$ operates only on
            the nonzero entries of $\mP$, and $\circ$ is the Hadamard (entrywise) product.
        \item The range of $\frac{d}{dt} \log \lambda_t$ is given by 
            \begin{gather}
            \inf_t \frac{d}{dt} \log \lambda_t 
                   = \liminf_{N \to \infty} \min_{\st \in \calP^N} \frac{1}{N} \log P(\st) \bydef \rho_{\min}
                   \label{eq:rho_min_def} \\
            \sup_t \frac{d}{dt} \log \lambda_t 
                    = \limsup_{N \to \infty} \max_{\st \in \calP^N} \frac{1}{N} \log P(\st) \bydef \rho_{\max}.
                    \label{eq:rho_max_def}
        \end{gather}
    \end{enumerate}
    \label{prop:loglmaxprops}
\end{proposition}
\begin{IEEEproof}
    See Appendix \ref{ap:loglmaxprops}.
\end{IEEEproof}

Now we can prove the following proposition:
\begin{proposition}
    \label{prop:LDP_rho}
$Q^1_N$ has a large deviations property with rate function
\begin{align*}
    I_1(\rho) & = \sup_t \set{t \rho - \log \lambda_t + \log \lambda_0} \\
&= \pathrate - s(\rho),
\end{align*}
where $s(\rho) \bydef \inf_t \Big\{ \log \lambda_t - t \rho \Big\}$.
\end{proposition}
\begin{IEEEproof}
    Since $\log \lambda_t$ is analytic, the limiting CGF $c(t)$ as defined in (\ref{eq:logP_cgf}) is differentiable, and
    the proposition follows from the G\"artner-Ellis theorem.
\end{IEEEproof}

To complete the large deviations analysis, we will need to use several properties of $s(\rho)$. One quantity that will
be important is the \emph{entropy rate} of $\mP$:
\begin{definition}
The entropy rate of an irreducible and aperiodic Markov chain $\mP$ is given by
\[
    H = -\sum_i \pi_i \sum_j p_{i,j} \log p_{i,j},
\]
where $\vec{\pi}$ is the unique stationary distribution. The entropy rate can be understood as the 
conditional entropy of the next state given the current state, averaged over the stationary distribution.
\end{definition}

This definition will be important in the following proposition:
\begin{proposition}
If $\mP$ is the transition matrix for a uniform random walk on a $K$-regular graph, then $s(\rho)$ is given by
\begin{equation}
    s(\rho) = \begin{cases}
        \log K & \text{if } \rho = -\log K \\
        -\infty & \text{if } \rho \neq - \log K.
    \end{cases}
  \label{eq:J_regular_graph}
\end{equation}

Otherwise, $s(\rho)$ satisfies the following properties:
\begin{enumerate}[label=(\arabic*),wide=0pt]
    \item $s: \R \to \R \bigcup \{-\infty\}$ is a concave function that is nonnegative on its effective domain,
        $[\rho_{\min}, \rho_{\max}]$, where $\rho_{\min}$ and $\rho_{\max}$ were defined in (\ref{eq:rho_min_def}) and
        (\ref{eq:rho_max_def}), respectively.
    \item $s(\rho)$ is continuous in $(\rho_{\min}, \rho_{\max})$, and continuous from above at $\rho_{\min}$ and $\rho_{\max}$.
    \item $s(\rho)$ is differentiable on $(\rho_{\min}, \rho_{\max})$. The function $s'(\rho)$ is one-to-one and $-s'(\rho)$
        is the inverse function of $\frac{d}{dt} \log \lambda_t$.
    \item \label{itm:JH} $s(-H) = H$ and $s'(-H) = -1$.
           Meanwhile, $s(\rho_0) = \log \lambda_0$ and $s'(\rho_0) = 0$, where
           $\rho_0 = \frac{(\va_0)^T  (\log \mP) \vb_0}{\va_0^T \vb_0}$.
  \end{enumerate}
\label{prop:Jprops}
\end{proposition}
\begin{IEEEproof}
    See Appendix \ref{ap:Jprops}.
\end{IEEEproof}

\begin{figure*} 
    \centering 
    \includegraphics[width=2.5in]{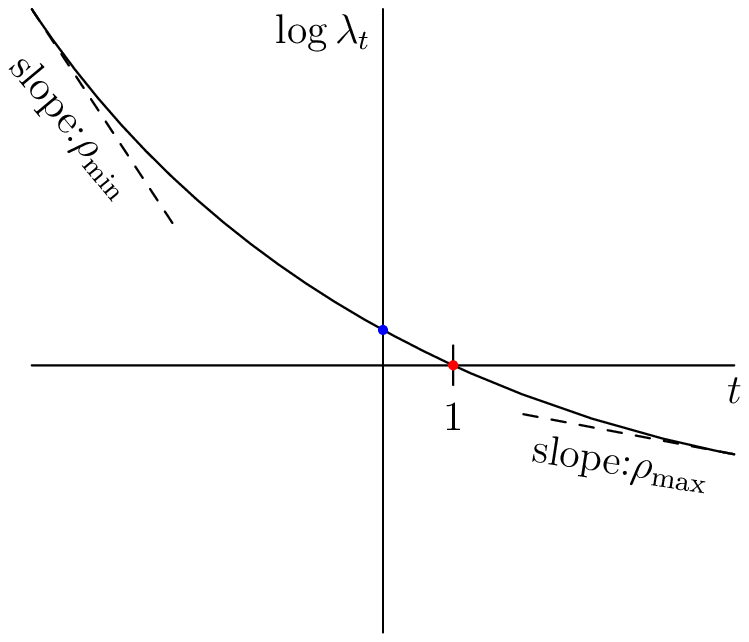}
    \hspace{0.25in}
    \includegraphics[width=3.5in]{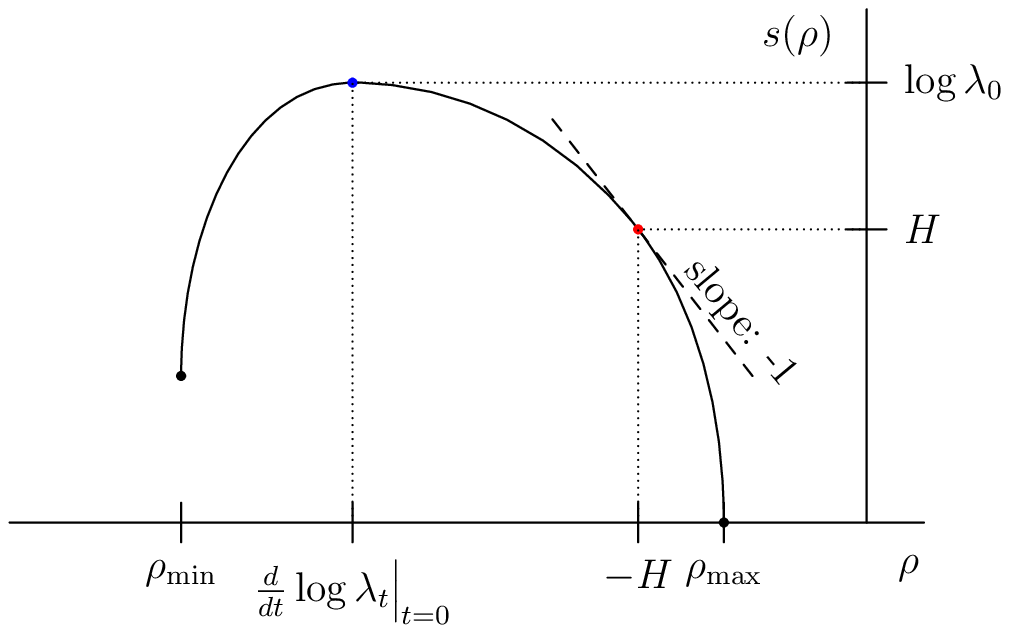}
    \caption{The basic properties of the functions $\log \lambda_t$ and $s(\rho)$ are illustrated here.
         $\log \lambda_t$ is a convex function (strictly convex except for a degenerate case); its value at $t=1$ is $0$, and it has limiting slopes $\rho_{\min}$ and $\rho_{\max}$.
         $s(\rho)$ is nonnegative and concave, takes the value $H$ at $\rho = -H$ (where the slope is $-1$), and is finite only on $[\rho_{\min}, \rho_{\max}]$.
         Its peak and the location thereof is determined by the value and slope,
         respectively, of $\log \lambda_t$ at $t=0$.
         (For the degenerate case of a uniform random walk on a $K$-regular graph, the curves look different:
         $\log \lambda_t$ is just a linear function $(1-t) \log K$, and $s(\rho)$ is only finite at a \emph{single point},
          $\rho = -\log K$, where $s(-\log K) = \log K$.)}
    \label{fig:loglt_and_Jrho}
\end{figure*}

We provide notional illustrations of $\log \lambda_t$ and $s(\rho)$ in the general case, based
on the properties described in Propositions \ref{prop:loglmaxprops} and \ref{prop:Jprops}, in Figure \ref{fig:loglt_and_Jrho}.

Now we can prove the large deviation property for the two-dimensional empirical measure $Q_N$ induced by the pairs $(\frac{1}{N} \log P(\st), \frac{1}{N} x_{\st})$:
\begin{theorem}
    \label{theorem:LDP_rhoxi}
With probability $1$, the empirical measure $Q_N$ satisfies the large deviation property with rate function
\begin{equation}
    I(\rho, \xi) = \begin{cases}
        I_1(\rho) + \frac{\xi^2}{2}, & \text{if } I_1(\rho) + \frac{\xi^2}{2} \leq \pathrate \\
        \infty, & \text{otherwise.}
    \end{cases}
    \label{eq:Irhoxi_def}
\end{equation}
\end{theorem}
\begin{IEEEproof}
    See Appendix \ref{ap:LDP_rhoxi}.
\end{IEEEproof}

\begin{remark}
    The microcanonical entropy density functions described in Section \ref{sec:nonrigorous}
    and the large deviation rate functions computed in this section are 
    closely related. Entropy density functions give the exponential growth rate
    for the number of states within some window; large deviation rate functions
    give the exponential decay rate for the probability of a uniformly chosen state
    in some window. Since the number of states in a window is equal to $\#\calP^N$
    times the probability under the empirical measure, we have that
    the microcanonical entropy density functions as illustrated in Figure \ref{fig:notional_J_and_srx} are given by:
    \[
        s(\rho) = \pathrate - I(\rho)
    \]
    and
    \begin{align}
        s(\rho, \xi) & = \pathrate - I(\rho, \xi) \nonumber \\
        & =  \begin{cases}
            s(\rho) - \frac{\xi^2}{2}, & \text{if }  |\xi| \leq \sqrt{2 s(\rho)} \\
             -\infty, & \text{otherwise}.
         \end{cases} \label{eq:srx_expression}
    \end{align}
\end{remark}

\subsection{The saddle point technique through Varadhan's lemma}
We can now compute the free energy density $\phiu(\beta)$ given in (\ref{eq:phiu_logPS_in_exponent}). We rewrite it in terms of the empirical measure as:
\begin{align}
\phiu(\beta) & = \lim_{N \to \infty} \frac{1}{N} \log(\#\calP^N)  
+ \lim_{N \to \infty} \frac{1}{N} \E \log \iint \exp(N [\beta \xi + \rho]) Q_N(d\rho, d\xi). \label{eq:phiu_as_integral_empirical_measure}
\end{align}
We have simply re-written the sum over all states as an integral over the discrete empirical measure induced by the states.
The first term is, as we know, $\pathrate$. The second term can be computed using Varadhan's lemma
\cite{ellis_entropy_2005}, a rigorous formulation of the Laplace principle (or the saddle point technique) applied to measures satisfying a large deviations property:
\begin{lemma}[{Varadhan's Lemma \cite[p. 51]{ellis_entropy_2005}}]
        Suppose a sequence $\{Q_N\}_{N=1}^\infty$ of probability measures on $\calX$ satisfies a large deviations property with
        rate function $I(\vx)$. Let $F: \calX \to \R$ be a continuous function that satisfies the tail condition 
        \begin{equation*}
            \lim_{L \to \infty} \limsup_{N \to \infty} \frac{1}{N} \log \int_{\vx: F(\vx) \geq L} \exp(N F(\vx)) Q_N(d\vx) = -\infty.
        \end{equation*}
        Then
        \begin{equation*}
            \lim_{N \to \infty} \frac{1}{N} \log \int_{\calX} \exp(N F(\vx)) Q_N(d\vx) = \sup_{\vx \in \calX} \Big\{ F(\vx) - I(\vx)\Big\}.
        \end{equation*}
\end{lemma}

We now have all the machinery in place to prove the main result:
\begin{theorem}
    \label{theorem:rigorous_saddlept}
     The free energy density is given by
     \begin{equation}
         \phiu(\beta) = \sup_{\rho, \xi } \Big\{s(\rho, \xi) + \beta \xi + \rho\Big\}, \label{eq:phiu_sup}
     \end{equation}
     where $s(\rho, \xi)$ is the microcanonical entropy density given in (\ref{eq:srx_expression}).
 \end{theorem}
\begin{IEEEproof}
To apply Varadhan's lemma, we need to show the tail condition 
\begin{equation*}
    \lim_{L \to \infty} \limsup_{N \to \infty} \frac{1}{N} \log \!\!\!\!\iint\limits_{(\rho,\xi):\beta \xi + \rho  \geq L} \!\!\!\!\!\exp(N [\beta \xi + \rho]) Q_N(d\rho, d\xi) = -\infty.
\end{equation*}
But this is simple. For all large enough $L$, the region $R = \{(\rho, \xi): \beta \xi + \rho \geq L\}$ has no
intersection with the support of $I(\rho, \xi)$, and thus it satisfies $Q_N(R) = 0$ with probability $1$.
Thus the tail condition holds, and Varadhan's lemma gives us that, almost surely,
\begin{align}
    \lim_{N \to \infty} \frac{1}{N} \log \sum_{\st \in \calP^N} \exp(\beta x_{\st} + \log P(\st)) & = \pathrate + \sup_{\rho, \xi} \Big\{\beta \xi + \rho - I(\rho, \xi)\Big\} \nonumber \\
    & = \sup_{\rho, \xi} \Big\{s(\rho, \xi) + \beta \xi + \rho\Big\}.
\end{align}

In general, almost sure convergence does not guarantee the convergence of the expectation. However, if a sequence of random variables is
\emph{uniformly integrable}, then almost sure convergence (indeed, merely convergence in probability) guarantees convergence in $L^1$, which is stronger than
convergence of the expectation. Uniform integrability is a sort of joint tail condition for a sequence of random variables. 
As it turns out, the sequence of random variables $\frac{1}{N} \log \Big( \sum_{\st} P(\st) \exp(\beta x_{\st}) \Big)$ is uniformly
integrable. Rather than belabor the point here, we will prove this fact (after formally defining uniform integrability)
in Appendix \ref{ap:uniform}.  This then immediately gives us the statement of the theorem.
\end{IEEEproof}

\subsection{Evaluating the bound}
\label{subsec:evaluating_phiu}

Now we are in a position to actually compute $\phiu(\beta)$, which will then give us a bound on the error exponent $\eta$.
We start with the degenerate case, which has a closed form expression:
\begin{proposition}
If $\mP$ is the transition matrix for a uniform random walk on a $K$-regular graph, then the error exponent satisfies
\begin{equation}
    \eta \geq
    \begin{cases}
        0, & \text{if } \beta \leq \sqrt{2 \log K} \\
    \frac{\beta^2}{2} - \beta \sqrt{2 \log K} + \log K, & \text{otherwise}.
    \end{cases}
    \label{eq:phiu_regular_graph}
\end{equation}
\label{prop:errexp_bound_regular}
\end{proposition}
\begin{IEEEproof}
    Combining (\ref{eq:J_regular_graph}), (\ref{eq:srx_expression}) and (\ref{eq:phiu_sup}), we have $\phiu(\beta) = \sup_{|\xi| \leq \sqrt{2 \log K}} \Big\{\beta \xi - \frac{\xi^2}{2}\Big\}$. 
    The supremum can be solved exactly; using the bound $\eta \geq \frac{\beta^2}{2} - \phiu(\beta)$ gives us (\ref{eq:phiu_regular_graph}).
\end{IEEEproof}

The general case is slightly more complicated. We have the following parametric representation (of which the
degenerate case expression given in Proposition \ref{prop:errexp_bound_regular} is a special case):
\begin{proposition}
For any irreducible and aperiodic Markov chain $\mP$, the error exponent bound is
\[
    \eta \geq \begin{cases}
        0, & \text{if } \beta \leq \sqrt{2 H} \\
        \chi(\beta), & \text{if } \beta \geq \sqrt{2 H},
    \end{cases}
\]
where $\chi(\beta)$ is a function that can be parametrized for $t \in (0,1]$ as:
\begin{equation}
\begin{gathered}
    \beta_t = \tfrac{\sqrt{2}}{t}\sqrt{\log \lambda_t - t \rho_t}, \\
    \chi(\beta_t) = \frac{1 - 2t}{t^2} \log \lambda_t - \frac{1-t}{t}\rho_t,
\label{eq:parametric_phiubeta}
\end{gathered}
\end{equation}
and $\rho_t = \frac{d}{dt} \log \lambda_t$ is given in (\ref{eq:loglt_deriv}). 
\label{prop:errexp_bound_general}
\end{proposition}
\begin{IEEEproof}
Since the function $s(\rho) - \frac{\xi^2}{2} + \beta \xi + \rho$ is concave and continuous on
the effective domain of $s(\cdot, \cdot)$, given by $\calA = \{(\rho, \xi) : |\xi| \leq \sqrt{2 s(\rho)}\}$, the supremum is achieved
at a point where $s'(\rho) = -1$ and $\xi = \beta$, if one exists in the interior of $\calA$; if not, then the supremum is achieved on the boundary of $\calA$.
See Figure \ref{fig:s_argmax} for an illustration.
From Proposition \ref{prop:Jprops}, we know that $s'(-H) = -1$ (the only such point), and $s(-H) = H$. 
So we get $\phiu(\beta) = H - \frac{\beta^2}{2} + \beta^2 - H = \frac{\beta^2}{2}$
so long as $\beta \leq \sqrt{2 H}$.

Otherwise, the supremum is achieved on the boundary, so $\xi = \sqrt{2 s(\rho)}$ and
\[
    \phiu(\beta) = \sup_{\rho \in [\rho_{\min}, \rho_{\max}]} \beta \sqrt{2 s(\rho)} + \rho.
\]
Since the function to be maximized is differentiable, the supremum occurs at the value of $\rho$ for which
\[
    \frac{\beta s'(\rho)}{\sqrt{2 s(\rho)}} + 1 = 0,
\]
if one exists; otherwise the supremum occurs at one of the endpoints $\rho_{\min}$ or $\rho_{\max}$. We
will show that such a point always exists. To see this, choose any $t \in (0,1]$. Based on the results
in Propositions \ref{prop:loglmaxprops} and \ref{prop:Jprops}, we know that for $\rho_t = \frac{d}{dt} \log \lambda_t$,
we have $s'(\rho_t) = -t$ and $s(\rho_t) = \log \lambda_t - t \rho_t$. This in turn gives us a value of $\beta$:
\[
    \beta_t = -\frac{\sqrt{2 s(\rho_t)}}{s'(\rho_t)} 
\]
and a corresponding value
\[
    \phiu(\beta_t) = -\frac{2 s(\rho_t)}{s'(\rho_t)} + \rho_t.
\]
Using these representations, we can compute $\beta_1$---since we know that $\frac{d}{dt} \log \lambda_t \Big|_{t = 1} = -H$, we have that $\beta_1 = \sqrt{2 H}$.
Meanwhile, $\lim_{t \to 0+} \beta_t = \infty$. This is because the numerator $\sqrt{2 s(\rho_0)} = \sqrt{2 \log \lambda_0} > 0$ by the Perron-Frobenius
theorem, so $s(\rho)$ is strictly positive in a neighborhood of $t=0$, while the denominator $s'(\rho_t)$ approaches $0$ from below.
From the intermediate value theorem, we can then achieve any value of $\beta$ in $[\sqrt{2 H}, \infty)$ by choosing some $t \in (0,1]$.
Thus we have a fully parametric representation, and substituting the known values of $s(\rho_t)$ and $s'(\rho_t)$ 
and applying the bound $\eta \geq \frac{\beta^2}{2} - \phiu(\beta)$ gives us the result.
\end{IEEEproof}

The bound given in Proposition~\ref{prop:errexp_bound_general} is equal to $0$ when the SNR is below a threshold: $\beta^2 \leq 2 H$. However, it is strictly positive for
SNR above the threshold. Thus, we can guarantee strong performance when the SNR is greater than twice the entropy rate of the Markov chain.
The entropy rate is smaller when the Markov structure is more restrictive; thus, the stronger our information about the dynamics of the process,
the stronger the performance of the detection. Furthermore, at very high SNR $\beta \gg 2 H$, we can use the parametric representation (\ref{eq:parametric_phiubeta}) to show that 
$\tfrac{\beta^2}{2} - O(\beta) \leq \eta \leq \tfrac{\beta^2}{2}$,
meaning the upper bound derived in Section \ref{subsec:bounds} becomes tight. This is to be expected; at very high SNR, the knowledge of the true state path is not necessary
to improve performance.

\subsection{Numerical Verification}
\label{sec:simulations}
From Lemma \ref{lemma:error_exponent_KL_rate}, which equates the error exponent to the Kullback-Leibler divergence rate, and 
Proposition \ref{prop:loglike_converges}, which says the normalized log likelihood ratio converges almost surely to $-\kappa = -\eta$,
and the fact that the log likelihood ratio can be computed efficiently, we have a simple Monte Carlo technique for estimating the true $\eta$. The only
caveat is to prevent numerical underflow through a suitable renormalization procedure.

\begin{figure*}[t]
    \centering
 \includegraphics[height=1.7in]{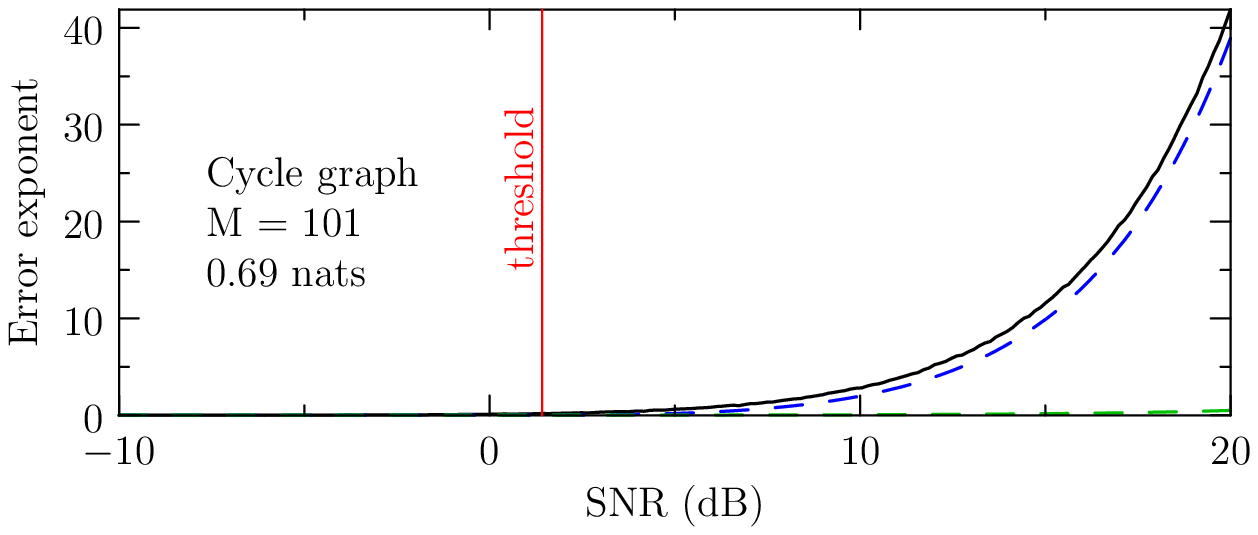} 

 \includegraphics[height=1.7in]{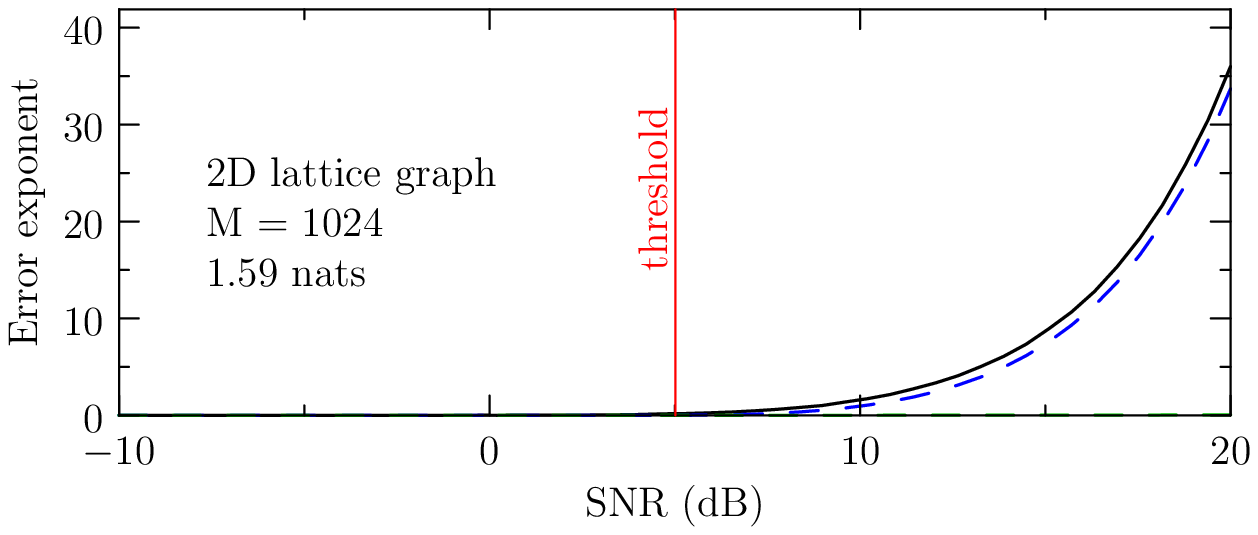}

  \includegraphics[height=1.7in]{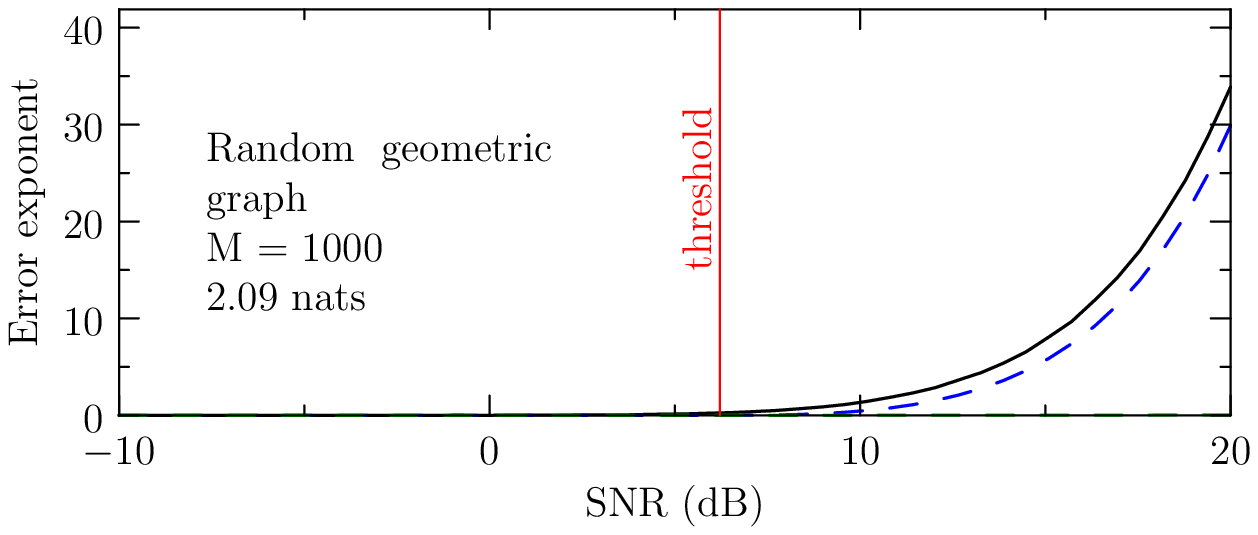}

 \includegraphics[height=1.7in]{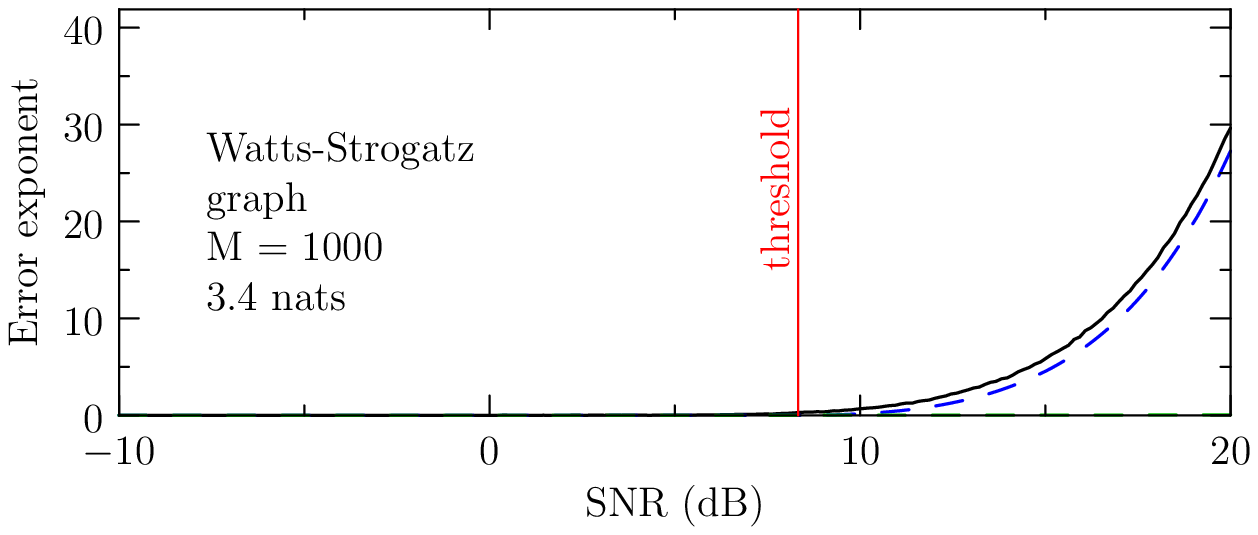}
 \caption{ Error exponent curves are plotted for random walks on four graphs, from top to bottom: 
             a  cycle graph with 101 vertices ($H$ = 0.693 nats),
             a $32 \times 32$ grid ($H$ = 1.58 nats),
             a random geometric graph with $1000$ vertices ($H$ = 2.09 nats),
             and a Watts-Strogatz small world graph \cite{watts_collective_1998} ($H$ = 3.41 nats). 
             The solid curve is the error exponent computed via Monte Carlo simulations.
             The green dashed curve is the sum-detector lower bound, which is barely nontrivial because $M$ is large.
             The blue dashed curve is our statistical physics-based analytic lower bound, computed using the parametric representation (\ref{eq:parametric_phiubeta}).
             The analytic threshold (SNR = $2 H$) is shown as well. At the same SNR level, the higher the entropy rate of the Markov chain, the worse the detector performance.
         }
    \label{fig:simulations}
\end{figure*}

We used this Monte Carlo technique to estimate the error exponents over a range of SNRs for several Markov chains.
In Figure \ref{fig:simulations} we compare the Monte Carlo simulations to the lower bound obtained
using the parametric representation
(\ref{eq:parametric_phiubeta}).

Although the phase transition appears only in the lower bound, the true error exponent curves appear to exhibit a smoothed version of the phase transition. 
Below the threshold the error exponent is quite small. It is bounded by the sum detector's error exponent of
$\frac{\beta^2}{2M}$, as we showed in Section \ref{subsec:bounds}. Of course, the sum detector completely ignores the structure of
the problem, and  when $M$ is large, this bound is practically $0$. Meanwhile, above the threshold the 
error exponent grows quickly with increasing SNR. Thus the simple test $\beta^2 \lessgtr 2 H$
suffices to determine whether one should expect good or bad detection performance.

\section{Conclusions}
\label{sec:conclusions}
In this paper, we studied the problem of detecting a random walk on a graph from spatiotemporal measurements corrupted by Gaussian noise. We modeled the problem
as a combinatorial hypothesis testing problem and studied the type-II error exponent of the optimal Neyman-Pearson detectors.
We proved the existence of the error exponent and the fact that it is equal
to the limiting Kullback-Leibler divergence rate between the two hypotheses. We showed how concepts from statistical physics could be used to analyze this
quantity, and rigorously proved a bound for the error exponent. Monte Carlo simulations show that, unlike the sum detector bound,
our bound fully captures the behavior of the error exponent.
In particular, the bound provides us with a simple test for whether to expect
strong or weak performance: if the SNR is greater than twice the entropy rate of the random walk, then detection will be easy.

\appendix

\subsection{Proof of Lemma \ref{lemma:error_exponent_KL_rate}}
\label{ap:error_exponent_KL_rate}
    
The proof is a rather straightforward generalization of the proof of the standard Chernoff-Stein lemma given in \cite{dembo_large_2009}. Consider the sequence of optimal detectors $\delta_N$, \emph{i.e.}, the Neyman-Pearson detector that choose $\calH_1$ if $\ell_N > \tau_N$ and $\calH_0$ otherwise,
    where $\tau_N$ is a sequence of thresholds chosen to satisfy the false alarm constraint $\PF \leq \epsilon$ for some fixed $\epsilon \in (0, 1)$. 
    The false alarm and miss detection probabilities are then given by
    \begin{align*}
        \PF^N & =  P_0( \ell_N > \tau_N) \\
        \intertext{and}
        \PM^N & = P_1( \ell_N < \tau_N),
    \end{align*}
    respectively.
    Note that we already have that $\liminf_{N \to \infty} \tau_N \geq -\kappa$;
    if that were not the case, then since $\ell_N  \to -\kappa$ in probability under $\calH_0$, we would have $\limsup_{N \to \infty} \PF^N = 1$, which would violate the false alarm constraint.

    Noting that $P_1(\mY^N) = \exp(N \ell_N) P_0(\mY^N)$, we can rewrite the miss detection probability as
    \begin{align}
        \PM^N & = \EE_1 \, \charfn(\ell_N < \tau_N) \nonumber \\
        & = \EE_0 \, \charfn(\ell_N < \tau_N) \exp(N \ell_N), \label{eq:PM_indicator_likelihood}
    \end{align}
    where $\charfn(\cdot)$ is the indicator function, since multiplying by $\exp(N \ell_N)$ converts the density $P_0(\cdot)$ to the density $P_1(\cdot)$. Choosing an arbitrary $\delta > 0$, we have
    \begin{align}
        & P_0(\ell_N \in [-\kappa - \delta, \tau_N]) \nonumber \\
        & = 1 - P_0(\ell_N < -\kappa - \delta) - P_0(\ell_N > \tau_N) \label{eq:nogoodnameforthis}\\
        & \geq 1 - P_0(\ell_N < -\kappa - \delta) - \epsilon,
    \end{align}
    since the final term in (\ref{eq:nogoodnameforthis}) is just the false alarm probability, which is constrained.
It then holds that
    \begin{align}
        \frac{1}{N} \log \PM^N & = \frac{1}{N} \log \EE_0\big[ \charfn(\ell_N < \tau_N) \exp(N \ell_N)\big] \nonumber\\
        & \geq \frac{1}{N} \log \EE_0 \big[\charfn(\ell_N \in [-\kappa - \delta, \tau_N]) \exp(N \ell_N)\big] \nonumber\\
        & \geq -\kappa - \delta + \frac{1}{N} \log P_0 \big( \ell \in [-\kappa-\delta, \tau_N] \big) \nonumber\\
        & \geq -\kappa - \delta + \frac{1}{N} \log\left[ 1 - \epsilon - \underbrace{P_0(\ell_N < - \kappa - \delta)}_{\longrightarrow 0}\right],\label{eq:liminf}
    \end{align}
    from which we can conclude that $\liminf_{N \to \infty} \frac{1}{N} \log \PM^N \geq -\kappa$, since $\delta$ can be made arbitrarily small and
    the last term on the right-hand side of \eref{liminf} vanishes.

    Now, instead, suppose we simply fix $\tau_N = -\kappa + \delta$ for every $N$. Then clearly $\PF^N \to 0$ because $\ell_N \to -\kappa$ in probability.
    Thus, eventually, $\PF^N < \epsilon$. Meanwhile, the maximum value of the quantity inside the integral in
    (\ref{eq:PM_indicator_likelihood}) is $\exp(N \tau_N)$, so we have that $\frac{1}{N} \log \PM^N \leq \tau_N = -\kappa + \delta$.
    So $\limsup_{N \to \infty} \frac{1}{N} \log \PM^N \leq -\kappa$, since again $\delta$ is arbitrary.
    
    We have shown the following: (1) any sequence of Neyman-Pearson detectors satisfying the false alarm constraint 
    $\PF^N < \epsilon$ satisfies 
    \[
    \liminf_{N \to \infty} \frac{1}{N} \log \PM^N \geq -\kappa,
    \]
    and (2) there exists a sequence of Neyman-Pearson detectors satisfying the false alarm constraint
    $\PF < \epsilon$ for which $\limsup_{N \to \infty} \frac{1}{N} \log \PM^N \leq -\kappa$.
    Thus for the optimal sequence of detectors, we have $\eta \bydef \lim_{N \to \infty} -\frac{1}{N}\log \PM^N = \kappa$. This holds for any $\epsilon \in (0, 1)$, so the proposition is proved.

\subsection{Proof of Proposition \ref{prop:loglmaxprops} [Properties of $\log \lambda_t$]}
\label{ap:loglmaxprops}
\begin{enumerate}[label=(\arabic*),wide=0pt,itemsep=2ex]
    \item $\hadpow{\mP}{t}$ is an irreducible nonnegative matrix for any $t$, just as $\mP$ is. Thus the
        Perron-Frobenius theorem tells us that $\lmax(\hadpow{\mP}{t})$ is a real, positive eigenvalue, so $\log
        \lambda_t$ is well-defined and finite.  Since $w^t$ is an analytic function of $t$ for any positive $w$ and the zero function is
        an analytic function, we have that every entry of $\hadpow{\mP}{t}$ is analytic in $t$. Standard
        perturbation-theoretic results \cite{kato_perturbation_2013} tell us that on any neighborhood in which a matrix
        function is analytic and an eigenvalue remains isolated from the rest of the spectrum (\emph{i.e.} has no multiplicity),
        it can be analytically continued to the rest of that neighborhood. Since $\lambda_t$ is the
        Perron-Frobenius eigenvalue, it is simple and thus isolated. Therefore, it is an analytic function of $t$ everywhere.
        Since it is positive, $\log \lambda_t$ is analytic as well. The convexity of $\log \lambda_t$ follows from a
        property of Hadamard powers proven by Horn and Johnson in \cite[p. 361]{horn_topics_1994}: for any nonnegative matrices
        $\mA$ and $\mB$ and $0 \leq \alpha \leq 1$, they showed that
        \begin{equation} 
            \lmax(\hadpow{\mA}{\alpha} \circ \hadpow{\mB}{1-\alpha}) \leq \lmax(\mA)^\alpha \lmax(\mB)^{1-\alpha}.
            \label{eq:hadamard_power_ineq}
        \end{equation} 
        Taking $\mA = \hadpow{\mP}{s}$ and $\mB = \hadpow{\mP}{t}$ for arbitrary $t > s > 0$, and using the fact that $\log$ is an increasing
        function, we have
        \begin{equation}
            \log \lmax(\hadpow{\mP}{\alpha s + (1-\alpha) t}) 
            \leq \alpha \log \lmax(\hadpow{\mP}{s}) + (1-\alpha) \log \lmax(\hadpow{\mP}{t}),
            \label{eq:convexity_of_loglmax}
        \end{equation}
        which by definition gives us the convexity of $\log \lambda_t$.
    \item Strict convexity means that the inequality in (\ref{eq:convexity_of_loglmax}) must be strict. Since $\log$ is
        strictly increasing, equality holds if and only if it holds in (\ref{eq:hadamard_power_ineq}), which for
        irreducible matrices holds if and only if there exists a positive scalar $\gamma$ and a positive diagonal matrix $\mD$ 
        such that $\gamma \mA = \mD^{-1} \mB \mD$ \cite[p. 361]{horn_topics_1994}. For our problem, then, equality holds if and only if there are some $t > s > 0$ such that for all $i,j$
        \[
            \gamma p_{ij}^s = \frac{d_j}{d_i} p_{ij}^t,
        \]
        for some positive constants $\gamma$ and $d_i, i=1, \ldots, M$. Thus either $p_{ij} = 0$ or
        \[
            p_{ij} = \gamma^{\frac{1}{t-s}} d_i^{\frac{1}{t-s}}  d_j^{\frac{1}{s-t}}.
        \]
        Summing over $j$ on both sides of this equation tells us that $d_i$ must be a constant. This means that all of the
        nonzero entries of $\mP$ must be constant. Since the row sums of $\mP$ must be one, this means that every
        row of $\mP$ having exactly $K$ nonzero entries equal to $\frac{1}{K}$ for some $K \leq M$ is the only
        situation in which strict convexity does not hold.

        So what exactly is $\log \lambda_t$ in that case? Consider the test vector $\vec{1}$: we have
        $\hadpow{\mP}{t} \vec{1} = K^{1-t} \vec{1}$. So the test vector is an eigenvector. The Perron-Frobenius
        theorem states that any positive eigenvector must correspond to the largest eigenvalue. Since $\vec{1}$
        has all positive entries, we have that $\lambda_t = K^{1-t}$, so $\log \lambda_t = (1-t) \log K$.

    \item As before, we use the perturbation results. In addition to an analytic eigenvalue function, in the case of a simple eigenvalue,
        there are analytic functions for the left- and right-eigenvectors. These can be normalized as desired. So we have analytic functions
        $\va_t$ and $\vb_t$ such that
        \begin{gather*}
            \va_t^T \hadpow{\mP}{t} = \lambda_t \va_t^T \\
            \hadpow{\mP}{t} \vb_t = \lambda_t \vb_t 
        \end{gather*}
        and normalized%
        \footnote{They are Perron-Frobenius eigenvectors, so they are positive and can never be orthogonal.}
        such that $\va_t^T \vb_t = 1$ 
        and $\va_t^T\vec{1} = 1$. We can write the largest eigenvalue function as $\lambda_t = \va_t^T \hadpow{\mP}{t} \vb_t$. Using the chain rule,
        we can compute the derivative 
        \begin{align}
            \lambda'_t & = (\va'_t)^T \hadpow{\mP}{t} \vb_t + \va_t^T \hadpow{\mP}{t} \vb'_t + \va_t^T ( (\log \mP) \circ \hadpow{\mP}{t}) \vb_t \nonumber \\
                      & = \lambda_t [ (\va'_t)^T \vb_t + \va_t^T \vb'_t] + \va_t^T ( (\log \mP) \circ \hadpow{\mP}{t}) \vb_t \nonumber \\
                      & = \va_t^T [ (\log \mP) \circ \hadpow{\mP}{t}] \vb_t, \label{eq:lambda_prime}
        \end{align}
        where in reaching (\ref{eq:lambda_prime}) we have used the fact that
        $\va_t^T \vb_t = 1$ and thus $(\va'_t)^T \vb_t + \va_t^T \vb'_t = 0$.
        Using the chain rule, we have that $\frac{d}{dt}\log \lambda_t = \frac{\lambda_t'}{\lambda_t}$, and the result follows.  Note that
        the normalization of the eigenvectors is irrelevant in the final expression because the normalization factors will cancel out in the numerator and denominator.

    \item  Since $\log \lambda_t$ is convex, its derivative is nondecreasing. 
           Thus 
           \begin{align*}
               \inf_t \frac{d}{dt} \log \lambda_t & = \lim_{t \to -\infty} \frac{d}{dt}\log \lambda_t \\
                                                  & = \lim_{t \to -\infty} \frac{\log \lambda_t}{t},
           \end{align*}
           where the finals step results from L'H\^opital's rule. By the same argument, we have
           \begin{equation*}
                \sup_t \frac{d}{dt} \log \lambda_t = \lim_{t\to +\infty} \frac{\log \lambda_t}{t}
           \end{equation*}
           Horn and Johnson show that these are equal to $\rho_{\min}$ and $\rho_{\max}$, respectively \cite[p. 367]{horn_topics_1994}.
\end{enumerate}

\subsection{Proof of Proposition \ref{prop:Jprops} [Properties of $s(\rho)$]}
\label{ap:Jprops}
As we stated in Proposition \ref{prop:loglmaxprops}, if $\mP$ is the transition matrix for a uniform random walk
on a regular graph, then $\log \lambda_t = (1-t) \log K$. If $\rho = -\log K$, then 
$\log \lambda_t - t \rho = \log K$, a constant, giving us $s(-\log K) = \log K$. For any other
$\rho$, the function $\log \lambda_t - t \rho$ is linear but not constant, so it is unbounded
and has an infimum of $-\infty$. Now consider the general case:
\begin{enumerate}[label=(\arabic*),wide=0pt,itemsep=2ex]
    \item Consider the function $-s(\rho) = \sup_t \set{t \rho - \log \lambda_t}$. This is
        the convex conjugate of $\log \lambda_t$.  A convex conjugate function is
        guaranteed to be a convex function with range $\R \bigcup  \{+\infty\}$,
        so $s(\rho)$ is a concave function with range $\R \bigcup \{-\infty\}$.
        Since $\log \lambda_t$ is strictly convex, the infimum 
        $\inf_t \log \lambda_t - t \rho$ that defines $s(\rho)$
        is achieved at no more than one point $t^*$ \cite{rockafellar_variational_1998}. 
        Since it is also differentiable, then if and only if the infimum is achieved at $t^*$, we have
        \[
            \frac{d}{dt} \log \lambda_t\Big|_{t^*} = \rho.
        \]
        If there is no such $t^*$, then $s(\rho) = -\infty$. This will be the case if $\rho < \rho_{\min}$ or $\rho > \rho_{\max}$.
        Suppose, however, that $\rho \in (\rho_{\min}, \rho_{\max})$. By the intermediate value theorem, there must be some $t^*$ for which
        $\frac{d}{dt} \log \lambda_t\Big|_{t^*} = \rho$. Then $s(\rho) = \log \lambda_{t^*} - t^* \rho$.
        It remains to prove nonnegativity.
        
        We use the following alternate expressions \cite[p. 367]{horn_topics_1994} for $\rho_{\min}$ and $\rho_{\max}$:
        \begin{align}
            \rho_{\min} & = \min_{\substack{\text{\rm self-avoiding loops}\\i_1,\ldots,i_L \\ 1 \leq L \leq M }} \frac{1}{L} \sum_{j=1}^{L} \log p_{i_j,i_{j+1}} \label{eq:rho_min_def_app}\\
                \rho_{\max} & = \max_{\substack{\text{\rm self-avoiding loops}\\i_1,\ldots,i_L \\ 1 \leq L \leq M }} \frac{1}{L} \sum_{j=1}^{L} \log p_{i_j,i_{j+1}}, \label{eq:rho_max_def_app}
             \end{align}
            where the suprema are over self-avoiding loops that obey the topology induced by the sparsity 
            of $\mP$, so each $i_1, \ldots, i_L$ is unique, $p_{i_j, i_{j+1}} \neq 0$, and we use the convention
            that $i_{L+1} = i_1$.)
            Let $i^*_1, \ldots, i^*_L$ be the self-avoiding loop achieving the maximum in (\ref{eq:rho_max_def_app}).
            Define the matrix $\mB$ as follows: every transition
            in the maximal loop is given the same value as in $\hadpow{\mP}{t}$ 
            (\emph{i.e.}, $[\mB]_{i^*_1,i^*_2} = p^t_{i^*_1,i^*_2}, \ldots, [\mB]_{i^*_L,i^*_1} = p^t_{i^*_L,i^*_1}$), and every other
            entry is set to $0$. On an elementwise basis, then, $\hadpow{\mP}{t} \geq \mB$. 
            If $L < M$, then $\mB$ is not irreducible, but the Perron-Frobenius theorem still
            guarantees us that it has a real eigenvalue $\lmax(\mB)$ equal to its spectral radius
            \cite{horn_matrix_2012}. It is not hard to verify that taking powers of
            $\mB$ eventually results in a constant multiple of a diagonal $0-1$ matrix: 
            \begin{align}
                \mB^{L} & = p_{i^*_1,i^*_2}^t \cdots p_{i^*_L,i^*+1}^t \mD \nonumber \\
                                      & = \exp( t L \rho_{\max}) \mD.
                                      \label{eq:BtotheL}
            \end{align}
            Here, the diagonal entries of $\mD$ associated with the indices $i^*_1, \ldots, i^*_L$ are
            $1$, and the others are all $0$.
            Now if we let $\vv$ be an eigenvector of $\mB$ associated with the eigenvalue
            $\lmax(\mB)$ whose only nonzero entries are those associated with the indices $i^*_1, \ldots, i^*_L$, we have
            \begin{equation}
                \mB^{L} \vv = \lmax(\mB)^L \vv.
            \end{equation}
            Combining this with (\ref{eq:BtotheL}), we obtain
            \[
            \lambda_{\max}(\mB) = \exp(t \rho_{\max})
            \]
            Since the Perron-Frobenius eigenvalue $\lambda_{\max}(\cdot)$
            is a monotonic function of the matrix entries \cite{horn_topics_1994},
            we have $\log \lambda_t \geq \log \lambda_{\max}(\mB^{(t)}) = t \rho_{\max}$ for every $t$. 
            Now since $\log \lambda_t - t \rho_{\max} \geq 0$ for all $t$, we must have $s(\rho_{\max}) \geq 0$.
            A similar argument shows that $s(\rho_{\min}) \geq 0$ as well.
            It then follows from the concavity of $s(\rho)$ that $s(\rho) \geq 0$ on $[\rho_{\min}, \rho_{\max}]$.

        \item Any proper convex function is continuous on the interior of its effective domain, so $-s(\rho)$ is
            continuous on $(\rho_{\min}, \rho_{\max})$, and thus $s(\rho)$ is as well.
            Since $-s(\rho)$ is the Legendre-Fenchel transform of $\log \lambda_t$, which is itself a convex function, it must be lower semicontinuous.  
            So $s(\rho)$ must be upper semicontinuous, and therefore it is continuous from above at $\rho_{\min}$ and $\rho_{\max}$.

        \item Since $\log \lambda_t$ is strictly convex (remember, we are not considering regular graphs here),
            there is at most one point that achieves the infimum $\inf_t \set{\lambda_t - t \rho}$
            that defines $s(\rho)$.
             We showed earlier that as long as $\rho \in (\rho_{\min}, \rho_{\max})$, there is exactly one such point. Another
            basic result in convex analysis \cite[Theorem 11.8]{rockafellar_variational_1998}  
            tells us that $s(\cdot)$ is then differentiable at $\rho$, and in particular $-s'(\rho) = t_\rho$, where $t_\rho$ is the argument of the minimum.
            Since $\log \lambda_t$ is differentiable, we also have that
            \[
                \frac{d}{dt} \log \lambda_t\Big|_{t = t_\rho} = \rho.
            \]
            Thus $-s'(\rho)$ is the inverse function of $\frac{d}{dt} \log \lambda_t$ as claimed. Since $\log \lambda_t$ is strictly convex, its derivative is one-to-one,
            and thus so is $s'(\rho)$.

        \item We know that $s(\rho) =  \log \lambda_{t_\rho} - t_\rho \rho$,
            where $t_\rho$ is the value of $t$ at which $\frac{d}{dt} \log \lambda_t = \rho$,
            if such a value exists, and $s'(\rho) = -t_{\rho}$; otherwise $s(\rho) = -\infty$.
            Using the expression for the derivative from the proof of Proposition \ref{prop:loglmaxprops}, and the fact
            that the left and right eigenvectors of $\mP$ are
            $\va_1 = \vec{\pi}$ and $\vb_1 = \vec{1}$, respectively, we have that 
            \begin{align*}
                \frac{d}{dt}\lambda_t\Big|_{t=1} & = \frac{\vec{\pi}^T[(\log \mP) \circ \mP] \vec{1}}{\vec{\pi}^T \mP \vec{1}} \\
                & = \sum_i \pi_i \sum_j p_{ij} \log p_{ij} \\
                & = -H.
            \end{align*}
            Meanwhile, $\lambda_1 = 1$, so $\frac{d}{dt} \log \lambda_t\Big|_{t=1} = -H$.
            So $s(-H) = \log 1 - 1\cdot(-H) = H$, and $s'(-H) = -1$.
            The same argument gives us $s(\rho_0)$ and $s'(\rho_0)$, only without the nicely-interpretable values. 
\end{enumerate}

\subsection{Proof of Theorem \ref{theorem:LDP_rhoxi}}
\label{ap:LDP_rhoxi}
To prove the statement of the theorem, we need to show that the upper bound
\begin{equation}
    \limsup_{N \to \infty} \frac{1}{N} \log Q_N(B) \leq -\inf_{(\rho, \xi) \in B} I(\rho, \xi)
    \label{eq:ldp_upper_bound_appendix}
\end{equation}
holds almost surely for every closed set $B \subset \R^2$, and the lower bound
\begin{equation}
    \liminf_{N \to \infty} \frac{1}{N} \log Q_N(U) \geq -\inf_{(\rho, \xi) \in U} I(\rho, \xi)
    \label{eq:ldp_lower_bound_appendix}
\end{equation}
holds almost surely for every open set $U \subset \R^2$.
We will use an argument parallel to Dorlas and Wedagedera's for the random energy model with an external field \cite{dorlas_large_2001}.
Let $\calA = \left\{ (\rho, \xi) : I_1(\rho) + \frac{\xi^2}{2} \leq \log \lambda_0 \right\}$ be the effective domain of $I(\cdot, \cdot)$,
\emph{i.e.} the set on which it is finite. It can also be written as $\calA = \left\{(\rho,\xi): |\xi| \leq \sqrt{2  s(\rho)} \right\}$. 
It is the union of hypograph of the function $\xi = \sqrt{ 2 s(\rho)}$
and its reflection over the $\rho$ axis. We know from Proposition \ref{prop:Jprops} that $s(\rho)$
is nonnegative and concave on $[\rho_{\min}, \rho_{\max}]$. Since $\sqrt{\;\cdot\;}$ is a concave and increasing function, we have that $\calA$ is a convex set. A notional illustration of $\calA$ was shown in Figure \ref{fig:s_argmax}.

We will be able to build up the result for general sets by studying the behavior of a few classes 
of primitive sets.  Consider a box $C = [\rho, \rho+\delta] \times [\xi, \xi+\delta]$ with sides of length $\delta$; suppose first that it
is entirely outside of $\calA$. By definition, $Q_N(C) = \frac{1}{\#\calP^N} \#\{\st : \frac{1}{N} \log P(\st) \in [\rho, \rho+\delta], \frac{1}{N} x_{\st} \in [\xi, \xi+\delta] \}$.
So $\#\calP^N Q_N(C)$ is a binomial random variable with parameters $\# \calP^N Q^1_N([\rho, \rho+\delta])$ and $\sqrt{\frac{N}{2 \pi}} \int_{\xi}^{\xi+\delta} \exp(-\frac{N x^2}{2}) dx$.
This means that
\begin{equation}
\E Q_N(C) = Q^1_N([\rho, \rho+\delta]) \sqrt{\frac{N}{2 \pi}} \int_{\xi}^{\xi+\delta} \exp(-\frac{N x^2}{2}) dx
\label{eq:E_QNC}
\end{equation}
and
\begin{multline*}
\var(Q_N(C)) = \frac{1}{\#\calP^N} Q_N^1([\rho, \rho+\delta])  
     \left(\sqrt{\frac{N}{2 \pi}} \int_{\xi}^{\xi+\delta} \exp(-\frac{N x^2}{2}) dx\right)  
     \left(1 - \sqrt{\frac{N}{2 \pi}} \int_{\xi}^{\xi+\delta} \exp(-\frac{N x^2}{2}) dx\right).
\end{multline*}
Now for any $\epsilon$, we can choose $N'$ large enough so that for every $N > N'$,
\begin{align}
     P( Q_N(C) > 0) 
    & = P( \#\calP^N Q_N(C) \geq 1) \label{eq:QNC_to_0:1} \\
    & \leq \E( \calP^N Q_N(C) ) \label{eq:QNC_to_0:2} \\
    & = \#\calP^N Q_N^1([\rho, \rho+\delta]) \sqrt{\frac{N}{2 \pi}} \int_{\xi}^{\xi+\delta} \exp(-\frac{N x^2}{2}) dx \label{eq:QNC_to_0:3} \\
    & \leq \exp\left( N\left[ \pathrate + \epsilon\right]\right)
       \exp\left(N \left[\epsilon -\inf_{r \in [\rho, \rho+\delta]} I_1(r)\right]\right)
       \delta \sqrt{\frac{N}{2 \pi}} \exp\left(N \left[-\inf_{x \in [\xi, \xi+\delta]} \frac{x^2}{2}\right]\right)    \nonumber  \\
    & = \exp\Bigg(N\bigg[-\inf_{\substack{r \in [\rho, \rho+\delta] \\ x \in [\xi, \xi+\delta]}} 
       \left(I_1(r) + \frac{x^2}{2}\right) + \pathrate + 2 \epsilon  \bigg] 
       + \log \delta + \frac{1}{2}\log \frac{N}{2 \pi}\Bigg) \rightarrow 0. \label{eq:QNC_to_0:5}
\end{align}
Here, (\ref{eq:QNC_to_0:1}) is because there is a discrete number of paths, (\ref{eq:QNC_to_0:2}) is the Markov inequality, and (\ref{eq:QNC_to_0:3}) is 
due to \eref{E_QNC}. This quantity converges to $0$ because the coefficient on $N$ in the exponent is guaranteed to be negative 
for small enough $\epsilon$ because
$C$ is entirely outside of $\calA$. We have merely proven convergence in probability, but since the probability goes to zero
exponentially fast, the Borel-Cantelli lemma tells us that with probability $1$, there is an $N'$ such that for every
$N > N'$, we have $Q_N(C) = 0$, so $\lim_{N \to \infty} \frac{1}{N} \log Q_N(C) = -\infty$ almost surely.

If instead we consider the half-planes $C = \{ (\rho, \xi) : \xi > \sqrt{2 \log \lambda_0} + 1\}$ or $C =  \{ (\rho, \xi) : \xi < -\sqrt{2 \log \lambda_0} - 1\}$,
we can use the same argument (replacing the Gaussian integrals in \eref{QNC_to_0:3} with standard Gaussian tail bounds, and using the fact that $C$ is outside of $\calA$) to show that $\lim_{N \to \infty} \frac{1}{N} \log Q_N(C) = -\infty$ almost surely
for these sets as well. The half planes $C = \{ (\rho, \xi): \rho > \rho_{\max}+1\}$ and $C = \{ (\rho, \xi) : \rho < \rho_{\min}-1\}$ also contain no states for large enough $N$ due to the
definitions of $\rho_{\min}$ and $\rho_{\max}$, so again $\lim_{N \to \infty} \frac{1}{N} \log Q_N(C) = -\infty$ almost surely.

Now, suppose we have a box $C = [\rho, \rho+\delta] \times [\xi, \xi+\delta]$, but this time it intersects the set $\calA$.
By Chebyshev's inequality we know that for any $\epsilon$, there is an $N'$ such that for $N > N'$,
\begin{align}
    & P\Big( |Q_N(C) - \E Q_N(C)| \geq k \E Q_N(C) \Big)  \nonumber \\
    & \leq \frac{1}{k^2} \frac{\var(Q_N(C))}{ (\E Q_N(C))^2 } \nonumber \\
    & \leq \frac{1}{k^2} \left(\#\calP^N \cdot Q_N^1([\rho, \rho+\delta]) \cdot \sqrt{\frac{N}{2 \pi}} \int_{\xi}^{\xi+\delta} \exp(-\frac{N x^2}{2}) dx\right)^{-1}\nonumber \\
    & \leq \frac{1}{k^2} \exp\Bigg(-N\Bigg[ \pathrate - \epsilon -\epsilon 
     - \inf_{r \in [\rho, \rho+\delta]} I_1(r) - \inf_{x \in [\xi, \xi+\delta]} \frac{x^2}{2} \Bigg]
     - \log \delta - \frac{1}{2} \log \left(\frac{N}{2 \pi}\right)  \Bigg). \nonumber
\end{align}
By choosing $\epsilon$ small enough, we can guarantee that this probability decays exponentially in $N$. Thus, by the Borel-Cantelli lemma
$\lim_{N \to \infty} \frac{Q_N(C)}{\E Q_N(C)} = 1$ with probability 1. Because $\log(\cdot)$ is continuous at $1$, this gives us
$\lim_{N \to \infty} \frac{1}{N} \log Q_N(C) = \lim_{N \to \infty} \frac{1}{N} \log \E Q_N(C)$.
Using (\ref{eq:E_QNC}), we can compute 
\begin{equation*}
    \lim_{N \to \infty} \frac{1}{N} \log Q_N(C) = -\inf_{(\rho,\xi) \in C} \Big\{I_0(r) + \frac{x^2}{2}\Big\} = -\inf_{(\rho,\xi) \in C} I(r,x),
\end{equation*}
almost surely. 

Using these primitives, we can prove the large deviation property directly. We start with the upper bound.
Suppose $B$ is a closed set entirely outside of the effective domain of $I(\cdot,\cdot)$, \emph{i.e.} $B \cap \calA = \emptyset$.
Let $d(B, \calA)$ be the distance between the set $B$ and $\calA$. Then if we choose some $\delta < d(B,\calA) / \sqrt{2}$, the set $B$
can be covered by a finite number of $\delta \times \delta$ boxes that are entirely outside of $\calA$  plus possibly one or more of the half-planes described above:
$B \subset \bigcup_{\ell=1}^L B_\ell$, where each $B_{\ell}$ is one of the primitives described above and $B_{\ell} \cap \calA = \emptyset$.
Then we have
\begin{align}
    \limsup_{N \to \infty} \frac{1}{N} \log Q_N(B) & \leq \limsup_{N \to \infty} \frac{1}{N}  \log \sum_{\ell = 1}^L Q_N(B_\ell) \nonumber \\
    & \leq \lim_{N \to \infty} \frac{1}{N} \log L + \limsup_{N \to \infty} \frac{1}{N} \log (\max_{\ell} Q_N(B_{\ell})) \nonumber \\
    & = -\infty, \label{eq:ldp_outside_A_is_minusinf}
\end{align}
almost surely, since the maximum is over just a finite number of sets.

Next suppose $B$ is a closed set that is not entirely outside the effective domain: $B \cap \calA \neq \emptyset$. Let $b = \inf_{(\rho, \xi) \in B} I(\rho, \xi)$.
Because $I$ is continuous inside $\calA$, for any $\epsilon$, we can choose a $\delta$ and cover $B$ with the primitive halfplanes and
a finite number of boxes of width $\delta$ such that, for each square (and each halfplane, trivially) $B_{\ell}$, we have $\inf_{(\rho, \xi) \in B_{\ell}} I(\rho, \xi) \geq b - \epsilon$. 
We have
\begin{align*}
    \limsup_{N \to \infty} \frac{1}{N} \log Q_N(B) & \leq \limsup_{N \to \infty} \frac{1}{N} \log \sum_{\ell = 1}^L Q_N(B_\ell) \\
    & \leq \lim_{N \to \infty} \frac{1}{N} \log L + \limsup_{N \to \infty} \frac{1}{N} \log (\max_{\ell} Q_N(B_{\ell})) \\
    & =\max_{\ell} \set{\limsup_{N \to \infty} \frac{1}{N} \log (Q_N(B_{\ell}))}  \\
    & \leq -(b - \epsilon).
\end{align*}
Since $\epsilon$ can be made arbitrarily small, we have
\begin{equation}
\limsup_{N \to \infty} \frac{1}{N} \log Q_N(B)  \leq -\inf_{(\rho, \xi) \in B} I(\rho, \xi). 
\end{equation}
Combining this with (\ref{eq:ldp_outside_A_is_minusinf}) gives us the large deviation upper bound for any closed set $B$.

Now we must prove the large deviations lower bound (\ref{eq:ldp_lower_bound_appendix}). Let $U$ be an open set. First, suppose $U \cap \calA = \emptyset$, meaning that the set is entirely outside the region.
Then the lower bound is trivial: it amounts to proving that $\liminf_{N \to \infty} \frac{1}{N} \log Q_N(U) \geq -\infty$, which is obviously true.
So let us assume that $U \cap \calA \neq \emptyset$.
For any $\epsilon > 0$, there is a square $C$ of width $\delta$ contained entirely within $U$ such that
$\inf_{(r, x) \in C} I(r,x) < \inf_{(r,x) \in U} I(r,x) + \epsilon$, by the following argument. First, note that the infimum must
be achieved on the interior of $\calA$%
\footnote{To see this, note that the boundary points of $\calA$ satisfy $\frac{\xi^2}{2} - s(\rho) = 0$,
in which case $I(\rho, \xi) = \log \lambda_0$ is the maximum possible value of $I$,
or either $\rho = \rho_{\min}, |\xi| \leq s(\rho_{\min})$ or $\rho = \rho_{\max}, |\xi| \leq s(\rho_{\max})$, in which case
the concavity of $s(\rho)$ tells us that we can decrease $I$ by moving into the interior of $\calA$.}%
.
If the infimum is achieved at a point $(\rho^*, \xi^*)$ on the interior of $U$, then we can easily just draw a box $C$ around it that is small enough to fit in $U$, and it
must have the same infimum. If on the other hand the infimum is achieved on the boundary of $U$, the continuity of $I(\rho, \xi)$ means that we can
choose a small open neighborhood around $(\rho^*, \xi^*)$ in which $I(\rho, \xi) < I(\rho^*, \xi^*) + \epsilon$. This neighborhood must
intersect with $U$ since it is centered on a boundary point, and the intersection must be an open set since both sets are open. Then we can choose a small
box $C$ that fits inside the intersection, and again we must have that $\inf_{(r, x) \in C} I(r,x) < \inf_{(r,x) \in U} I(r,x) + \epsilon$.

Using our result for boxes, we have
\begin{align*}
    \liminf_{N \to \infty}\frac{1}{N} \log Q_N(U) & \geq \liminf_{N \to \infty} \frac{1}{N} \log Q_N(C) \\
    & = -\inf_{(r,x) \in C} I(r,x) \\
    & > -\inf_{(r,x) \in U} I(r,x) - \epsilon,
\end{align*}
almost surely.
Since this holds for any $\epsilon$, the lower bound is proved.

\subsection{Uniform integrability of the free energy density}
\label{ap:uniform}
In this appendix, we show that the sequence of random variables
\begin{equation}\label{eq:rv_xn}
    X_N \bydef \frac{1}{N} \log \Big(\sum_{\st \in \calP^N} P(\st) \exp(\beta x_{\st}) \Big), \qquad N = 1, 2, \ldots
\end{equation}
is \emph{uniformly integrable}. Our arguments will closely follow those in Olivieri and Picco \cite{olivieri_existence_1984}, who showed that the free energy density of the standard random energy model \cite{derrida_random-energy_1981} is uniformly integrable. We start by recalling the definition of uniform integrability:

\begin{definition}
    A sequence of random variables $\set{X_N}_{N \ge 1}$ is uniformly integrable if
    \begin{equation}
        \lim_{\alpha \to \infty} \sup_{N > N_0} \E \Big(\mathds{1}_{\abs{X_N} \geq \alpha} \abs{X_N} \Big) = 0, \label{eq:uniform}
    \end{equation}
for some $N_0 > 0$.
\end{definition}

To proceed, we first note that, by the definition of $\rho_{\min}$ in \eref{rho_min_def}, there exists some $N_0$ such that $P(\st) > \text{exp}(2 N\rho_{\min})$ for all $N \ge N_0$. (Note that $\rho_{\min}$ is negative, so $2 \rho_{\min}$ is actually less than $\rho_{\min}$.) Using this inequality and the fact that $\sum_{\st \in \calP^N} P(\st) = 1$, we can bound the random variable $X_N$ in \eref{rv_xn} on both sides as 
\begin{equation}
    2\rho_{\min} + \frac{\beta}{N} \max_{\st \in \calP^N} x_{\st} \leq X_N \leq \frac{\beta}{N} \max_{\st \in \calP^N} x_{\st}. \label{eq:XNbounds}
\end{equation}
Now take any $\alpha \geq 1$. We can split the expectation in (\ref{eq:uniform}) into two parts and apply (\ref{eq:XNbounds}):
\begin{align}
\E \Big(\charfn_{|X_N| \geq \alpha} |X_N| \Big) & = \E \Big(\mathds{1}_{X_N \geq \alpha} X_N \Big) + \E \Big(\mathds{1}_{X_N \leq -\alpha} (-X_N) \Big) \nonumber \\
& \leq \E \Big(\charfn_{\frac{\beta}{N} \max_{\st} x_{\st} \geq \alpha} [\frac{\beta}{N} \max_{\st} x_{\st}]\Big) \nonumber
      + \E \Big(\charfn_{2\rho_{\min} +\frac{\beta}{N} \max_{\st} x_{\st} \leq -\alpha} [-2\rho_{\min} - \frac{\beta}{N} \max_{\st} x_{\st}]\Big) \\
      & \leq \sum_{K = 1}^\infty \alpha (K+1) P( \max_{\st}  x_{\st} \geq \alpha K N/\beta)  \nonumber \\
      & \quad\qquad + \sum_{K = 1}^\infty \alpha (K+1) P(\max_{\st} x_{\st} \leq -\alpha K N/\beta - 2 \rho_{\min}N/\beta), \label{eq:charfnexpbnd}
\end{align}
where to reach \eref{charfnexpbnd} we have simply decomposed the integrals corresponding to the expectations into a sum of integrals from $K\alpha$ to $(K+1)\alpha$ for $K = 1, 2, \ldots$, and bounded each one.

Let us consider the first probability expression in (\ref{eq:charfnexpbnd}). Defining $\Phi(\cdot)$ as the standard Gaussian cumulative
distribution function, and exploiting the fact that the ensemble $\{x_{\st}\}$ is i.i.d., we have
\begin{align}
P(\max_{\st}  x_{\st} \geq \alpha K N/\beta)\nonumber
        & = 1 - P\Big( x_{\st} \leq \alpha K N/\beta, \; \text{for all } \st\in\calP^N\Big) \nonumber\\
        & = 1 - \Big( 1 - \Phi(-\alpha K \sqrt{N}/\beta)  \Big)^{\# \calP^N} \nonumber\\
        & \leq \#\calP^N \exp\Big( \frac{-\alpha^2 K^2 N}{2\beta^2} \Big),\label{eq:tail_bnd}
\end{align}
where in reaching \eref{tail_bnd} we have used the inequality $(1-x)^K \ge 1 - Kx$ for any positive integer $K$ and any $x < 1$, and applied the standard Gaussian tail bound $\Phi(-t) \le \exp(-t^2/2)$ for $t >0$ (see, \emph{e.g.}, \cite[p. 445]{talagrand_mean_2010}). Recall that $\#\calP^N = \exp( N \log \lambda_0  + o(N))$. Thus, for all sufficiently large $N$ and sufficiently large $\alpha$, we can bound the first term on the right-hand side of \eref{charfnexpbnd} as
\begin{align*}
\sum_{K = 1}^\infty \alpha (K+1) P( \max_{\st}  x_{\st} \geq \alpha K N/\beta) &\le \sum_{K = 1}^\infty \alpha (K+1) \exp\Big(2N\log \lambda_0 - \frac{\alpha^2 K^2 N}{2\beta^2}\Big)\\
&\le \alpha \lambda_0^2 \sum_{K = 1}^\infty (K+1) \exp\Big(- \frac{\alpha^2 K^2}{2\beta^2}\Big),
\end{align*}
which converges to zero as $\alpha \to \infty$. Similar bounds allow us to reach the same conclusion for the second term on the right-hand side of \eref{charfnexpbnd}. It then follows that the uniform integrability condition \eref{uniform} holds for the sequence of random variables in \eref{rv_xn} corresponding to the free energy density.

\providecommand{\url}[1]{}
\renewcommand{\url}[1]{}
\bibliographystyle{IEEEtran}
\bibliography{nourl,strings,randomwalk}

\end{document}